%% file: main.tex
\documentclass[%
 reprint,
superscriptaddress, 
 amsmath,amssymb,
 aps,
]{revtex4-2}

\usepackage{graphicx}
\usepackage{dcolumn}
\usepackage{bm}
\usepackage{siunitx}
\usepackage{import}
\usepackage{lipsum}  
\DeclareMathOperator{\sinc}{sinc}
\newcommand*{\hatH}{\hat{\mathcal{H}}}
\usepackage{mathtools}
\usepackage{braket}
\usepackage{cancel}

\begin{document}


\input{Heading}
\input{Abstract}

\maketitle
\input{main_body/Introduction}
\input{main_body/Section2}
\input{main_body/Section3}
\input{main_body/Section4}
\input{main_body/Section5}
\input{Acknowledgments}
\appendix
\renewcommand\thefigure{A\arabic{figure}}
\renewcommand\thetable{A\Roman{table}}
\setcounter{figure}{0}
\setcounter{table}{0}
\input{appendix_body/Supplementary_A}
\input{appendix_body/Supplementary_B}
\input{appendix_body/Supplementary_C}
\input{appendix_body/Supplementary_D}
\input{appendix_body/Supplementary_E}
\input{appendix_body/Supplementary_F}

\input{appendix_body/Supplementary_G}

\bibliography{complete_paper_references_example}
\end{document}

%% file: Heading.tex
\title{Fast multiplexed superconducting qubit readout with intrinsic Purcell filtering} 
\author{Peter A. Spring}
\email{peter.spring@riken.jp}
\affiliation{%
RIKEN Center for Quantum Computing (RQC), Wako, Saitama 351-0198, Japan
}
\author{Luka Milanovic}
\affiliation{%
RIKEN Center for Quantum Computing (RQC), Wako, Saitama 351-0198, Japan
}
\affiliation{%
Department of Physics, ETH Zürich, Otto-Stern-Weg 1, 8093 Zürich, Switzerland
}
\author{Yoshiki Sunada}
\affiliation{%
RIKEN Center for Quantum Computing (RQC), Wako, Saitama 351-0198, Japan
}
\affiliation{%
 QCD Labs, QTF Centre of Excellence, Department of Applied Physics, Aalto University, P.O.\ Box 13500, FIN-00076 Aalto, Finland
}
\author{Shiyu Wang}
\affiliation{%
RIKEN Center for Quantum Computing (RQC), Wako, Saitama 351-0198, Japan
}
\author{Arjan F. van Loo}
\affiliation{%
RIKEN Center for Quantum Computing (RQC), Wako, Saitama 351-0198, Japan
}
\affiliation{Department of Applied Physics, Graduate School of Engineering, The University of Tokyo, Bunkyo-ku, Tokyo 113-8656, Japan}
\author{Shuhei Tamate}
\affiliation{%
RIKEN Center for Quantum Computing (RQC), Wako, Saitama 351-0198, Japan
}
\author{Yasunobu Nakamura}%
\affiliation{%
RIKEN Center for Quantum Computing (RQC), Wako, Saitama 351-0198, Japan
}
\affiliation{Department of Applied Physics, Graduate School of Engineering, The University of Tokyo, Bunkyo-ku, Tokyo 113-8656, Japan}

%

\date{\today}

%% file: Abstract.tex
\begin{abstract}
Fast and accurate qubit measurement remains a critical challenge on the path to fault-tolerant quantum computing. In superconducting quantum circuits, fast qubit measurement has been achieved using a dispersively coupled resonator with a large external linewidth. This necessitates the use of a Purcell filter that protects the qubit from relaxation through the readout channel. Here we show that a readout resonator and filter resonator, coupled to each other both capacitively and inductively, can produce a compact notch-filter circuit that effectively eliminates the Purcell decay channel through destructive interference. By utilizing linewidths as large as \SI{42}{\mega\hertz}, we perform $56$-\si{\nano\second} simultaneous readout of four qubits and benchmark an average assignment fidelity of $99.77\%$, with the highest qubit assignment fidelity exceeding $99.9\%$. These results demonstrate a significant advancement in speed and fidelity for multiplexed superconducting qubit readout.
\end{abstract}

%% file: main_body/Introduction.tex
\section{Introduction}
Quantum computing relies on the accurate measurement of qubit states. In superconducting quantum circuits, measurement now constitutes a major component of the error budget for algorithms requiring detection and feedback. This is significant for the exceution of quantum error correction codes such as the surface code~\cite{acharya2024quantum, google2023suppressing, zhao2022realization, krinner2022realizing, marques2022logical, Fowler2012}, which require the measurement of ancilla qubits at the end of every error-correction cycle. It is equally important in measurement-based protocols for preparing entangled states~\cite{tantivasadakarn2024long, chen2023realizing, zhu2023nishimori, lu2022measurement, raussendorf2003measurement} and performing long-range entangling gates~\cite{baumer2024efficient}, where unitary qubit operations are replaced by mid-circuit measurements and dynamical feedback.
\\
\indent Fast superconducting qubit readout has been achieved by dispersively coupling qubits to readout resonators that possess large external linewidths~\cite{jeffrey2014fast, Walter2017, Heinsoo2018, sunada2022fast, swiadek2023enhancing}. By further inserting a dedicated filter resonator between each readout resonator and the readout line~\cite{Heinsoo2018}, it is possible to suppress the Purcell relaxation channel and to reduce measurement crosstalk simultaneously.
\\
\indent However, as readout resonators with very large effective external linewidths are explored~\cite{sunada2024photon} and as energy relaxation times of superconducting qubits continue to improve well beyond \SI{100}{\micro\second}~\cite{place2021new, spring2022high, wang2022towards, deng2023titanium, kono2023mechanically, biznarova2023mitigation, somoroff2023millisecond}, the Purcell filtering in circuits employing filter resonators may be insufficient to ensure that the Purcell relaxation channel remains adequately suppressed. This motivates the design of filter resonators with an auxiliary Purcell filtering mechanism.
\\
\indent An attractive approach to improve Purcell filtering is to engineer an `intrinsic' filter that leverages properties of the circuit to enhance the filtering without the need for additional circuit components. Example intrinsic filtering mechanisms have involved exploiting symmetry in the qubit~\cite{roy2017implementation, pfeiffer2023efficient, Hazra2024benchmarking}, making use of a stray
capacitance in the readout circuit~\cite{Bronn2015intrinsic}, and utilizing the distributed nature of the readout resonator~\cite{sunada2022fast} and the readout feedline~\cite{yen2024interferometric}.
\\
\indent Here, we propose and experimentally demonstrate a readout-resonator and dedicated-filter configuration featuring an intrinsic notch-type filter. This is achieved by coupling two quarter-wavelength ($\lambda/4$) coplanar waveguide (CPW) resonators together through a multiconductor transmission line (MTL)~\cite{paul2007analysis}, which is formed by bringing a segment of the resonator lines into close proximity. The capacitive and inductive interactions in the MTL destructively interfere, resulting in a notch frequency where signals cannot propagate through the resonators. We experimentally demonstrate that this filtering mechanism can effectively eliminate the Purcell relaxation channel when the notch is tuned to the qubit frequency. By utilizing readout modes with large effective linewidths, we then perform frequency-multiplexed readout of four qubits using a sub-60-\si{\nano\second} measurement pulse and integration window, with the highest qubit assignment fidelity exceeding $99.9\si{\percent}$.
\\ 
\indent The paper is organized as follows: First, a distributed-circuit model and an equivalent lumped-element circuit model for the coupled resonators are presented and utilized to find symbolic expressions for the notch frequency and the coupling strength between the resonators. Next, the notch filtering is experimentally demonstrated and the predictions of the symbolic formulas are verified. Finally, fast four-to-one multiplexed readout is characterized in a device featuring readout and filter resonators with intrinsic notch filtering.

%% file: main_body/Section2.tex
\section{Model}
\begin{figure}
\includegraphics[width=0.475\textwidth,height=0.49\textheight,keepaspectratio]{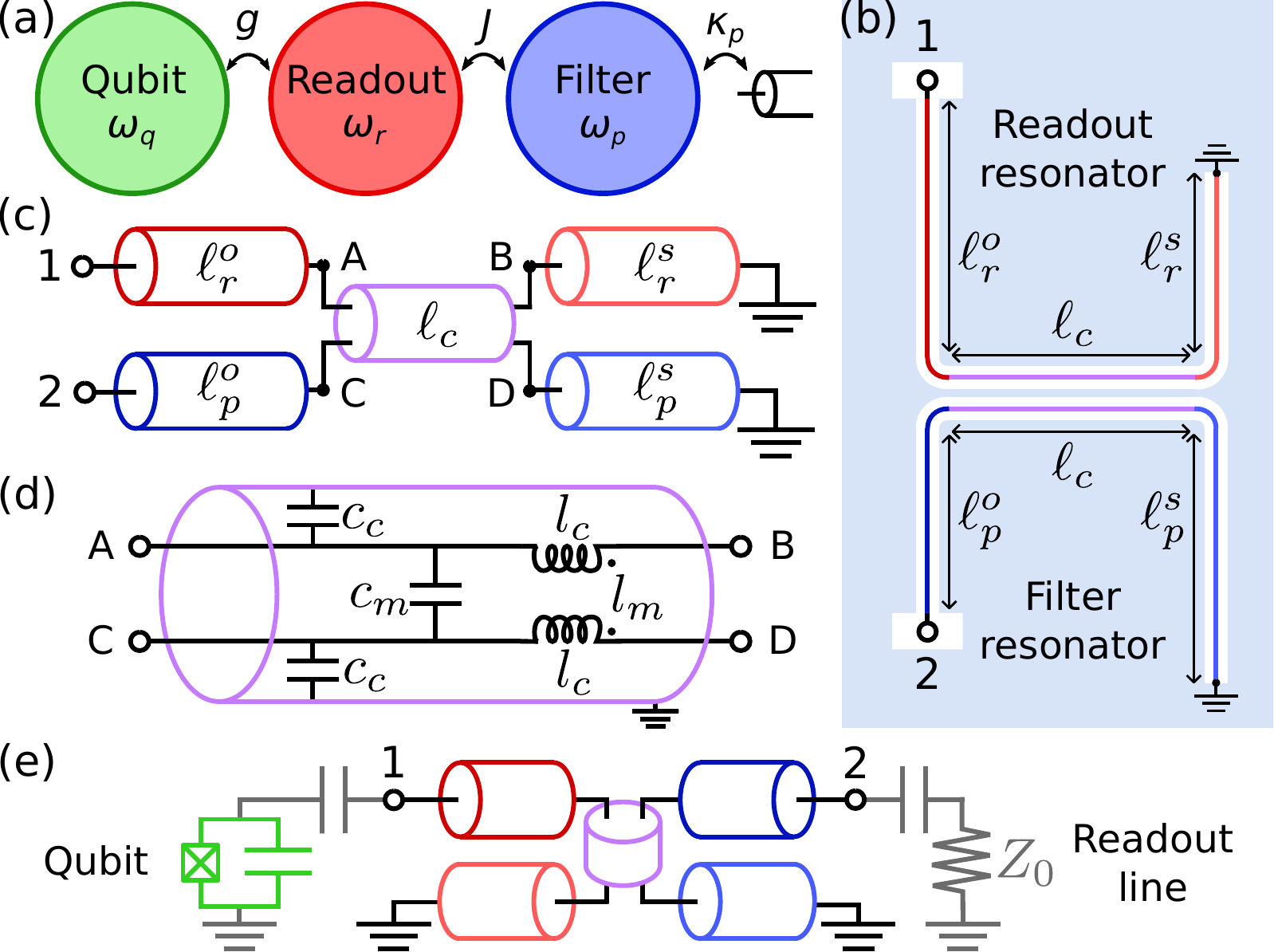}
\caption{\label{fig: schematics} Readout resonator and filter resonator with auxiliary notch filtering. (a)~General arrangement of dispersive readout with a readout resonator and dedicated filter. (b)~CPW configuration used in this work to construct a readout resonator and filter resonator with auxiliary notch filtering. (c)~Distributed-circuit-model representation. The coupled section is treated as a MTL with two coupled lines. (d)~Model of the MTL with the per-length parameters defined. (e)~Depiction of a qubit coupled to the readout line through the distributed readout- and filter-resonator circuits.}
\end{figure}

The general configuration of dispersive readout with a dedicated filter resonator is shown in  Fig.~\ref{fig: schematics}(a). A qubit is dispersively coupled to a readout resonator with coupling strength $g$, which in turn is coupled to a filter resonator with coupling strength $J$. The filter resonator is then coupled to a readout line with an external linewidth of $\kappa_p$. Generally, it is advantageous to minimize the detuning between the readout resonator and filter resonator such that $|\omega_r - \omega_p| \ll J $ in order to maximize the rate at which incident readout photons acquire information about the qubit state~\cite{swiadek2023enhancing, valles2023post}.
\\
\indent Figure~\ref{fig: schematics}(b) introduces the readout- and filter-resonator configuration used here to realize intrinsic notch filtering. Two $\lambda/4$ CPW resonators having characteristic impedance $Z_0$ and phase velocity $v$ are coupled together through a length-$\ell_c$ section where the lines are brought into close proximity. The total lengths of the readout and filter resonators are $\ell_{r} = \ell_{r}^o + \ell_c + \ell_{r}^s$ and $\ell_{p} = \ell_{p}^o + \ell_c + \ell_{p}^s$, respectively, and their fundamental $\lambda/4$ resonances are at $\omega_{r(p)} = \pi v/ \left(2\ell_{r(p)}\right)$. As depicted in the figure, $\ell_{r(p)}^{o}$($\ell_{r(p)}^{s}$) denotes the lengths of line spanning between the open(shorted) end of each resonator and the coupled section.
\\
\indent Figure~\ref{fig: schematics}(c) presents the distributed circuit used to model the coupled resonators. The length-$\ell_c$ region where the coupling occurs is treated as an MTL with parameters defined in Fig.~\ref{fig: schematics}(d). The two lines of the MTL are coupled together by capacitance and inductance per length, $c_m$ and $l_m$, respectively. Both lines are symmetric with the same capacitance-to-ground and self-inductance per length, $c_c$ and $l_c$, respectively.
\begin{figure}
\includegraphics[width=0.39\textwidth,height=0.49\textheight,keepaspectratio]{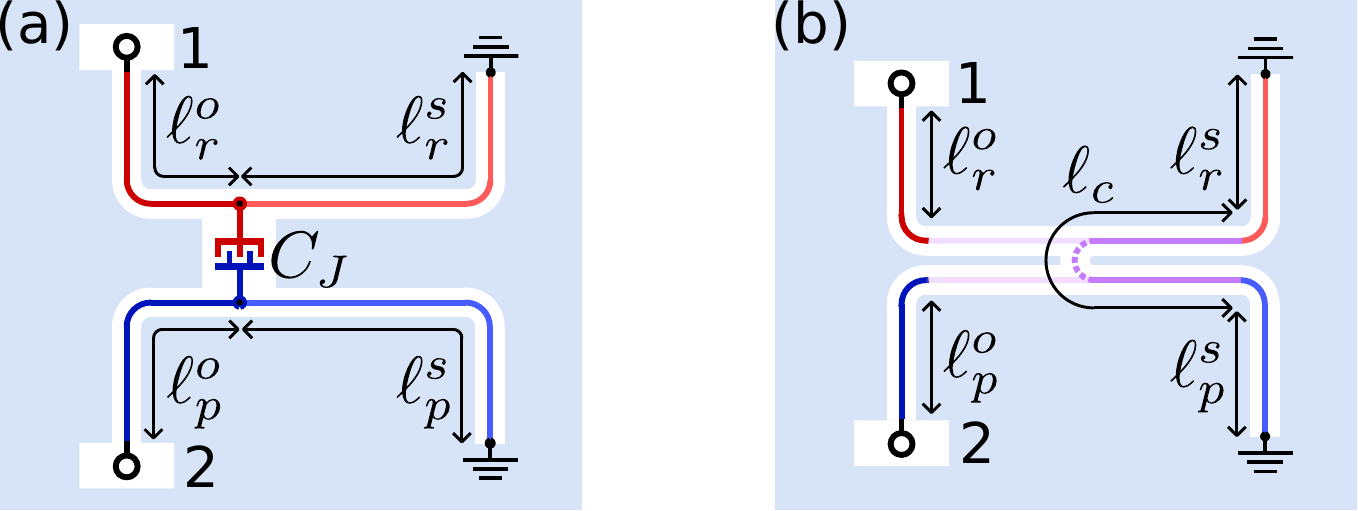}
\caption{\label{fig:Useful coupled resonator diagrams} Schematics of coupled $\lambda/4$ CPW resonators. (a) Diagram of lumped-element capacitive coupling between a pair of $\lambda/4$ resonators. (b) Geometric interpretation of the intrinsic notch that arises between a pair of MTL-coupled $\lambda/4$ resonators. The notch occurs at the anti-resonance frequency of the effective $\lambda/2$ resonator formed between the two resonators having a total length of $\ell_r^s + \ell_c + \ell_p^s$.}
\end{figure}
\begin{figure*}
\includegraphics[width=0.99\textwidth,height=0.49\textheight,keepaspectratio]{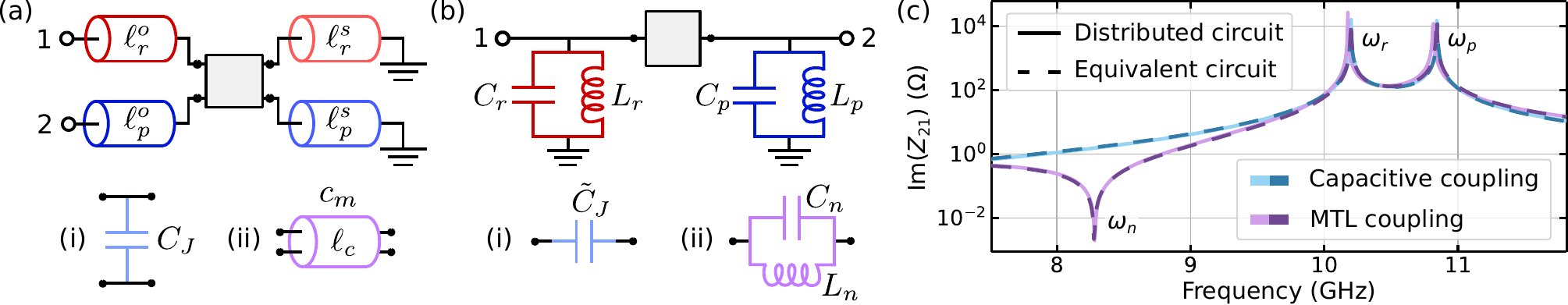}
\caption{\label{fig:equivalent circuits} (a)~Distributed-circuit model for the coupled $\lambda/4$ resonators. The coupling element~(gray square) represents (i)~a lumped capacitor or (ii)~a MTL. (b)~Equivalent lumped-element-circuit models. The coupling element is represented as (i)~a lumped capacitor in the case of direct capacitive coupling or (ii)~an LC resonator in the case of MTL coupling. (c)~Imaginary part of the transfer impedance $Z_{21}$ of the capacitive and MTL coupling circuits for the set of parameters given in Table~\ref{tab:resonator dimensions - appendix} in Appendix~\ref{appendix:Deriving equivalent circuits and Interaction Hamiltonian}. Solid curves are the frequency dependence obtained from the distributed-circuit models, and dashed curves are from the equivalent-circuit models. The coupling strength between the resonators is $J/2\pi=\SI{30}{\mega\hertz}$ in both cases.}
\end{figure*}
\\
\indent
By making appropriate assumptions about the MTL section, a symbolic expression for the transfer impedance $Z_{21}(\omega)\equiv V_2(\omega)/I_1(\omega)|_{I_2=0}$ between the two ports at the open ends of both resonators is derived [see Eq.~\eqref{eq:Z21 weak coupling general result} in Appendix~\ref{appendix:Transfer impedance solution}]. At frequencies where $Z_{21}=0$, the qubit in Fig.~\ref{fig: schematics}(e) is decoupled from the readout line because no amount of excitation applied at port 1 of the readout resonator can induce a voltage at port 2 of the filter resonator. Using this general solution, we proceed to analyze the behavior of the two resonators in the cases they are capacitively and MTL coupled. The capacitive coupling case serves as a reference in order to quantify the enhancement in Purcell filtering achieved through MTL coupling.
\subsection{Capacitive coupling}
Capacitive interaction between two $\lambda/4$ CPW resonators, i.e., a readout resonator and filter resonator, as depicted in Fig.~\ref{fig:Useful coupled resonator diagrams}(a) has been widely used for superconducting qubit readout~\cite{Heinsoo2018, krinner2022realizing, marques2022logical, swiadek2023enhancing, levine2023demonstrating}. 
A lumped-element capacitor represents a special case of the MTL depicted in Fig.~\ref{fig: schematics}(d), corresponding to the limits $c_m \ell_c \to C_J$ and $l_m, \ell_c \to 0$. In this case, the solution to $Z_{21}(\omega)$ [Eq.~\eqref{eq:Z21 for direct capacitive coupling} in Appendix~\ref{appendix:Transfer impedance solution}] has its first zero at a frequency greater than or equal to $\textrm{min}\left(2\omega_r, 2\omega_p\right)$, confirming that it is not possible to engineer a notch filter in the relevant frequency range for Purcell filtering by using a purely capacitive coupling.
\subsection{MTL coupling}\indent Coupled CPWs on a substrate with permittivity $\epsilon_r$ behave to a good approximation as if they were embedded in a homogeneous medium with an effective permittivity of $\epsilon_{\textrm{eff}}=(1+\epsilon_r)/2$~\cite{ghione1994efficient}. By applying this result to the MTL that couples the resonators, the solution to $Z_{21}(\omega)$ becomes (see Appendix~\ref{appendix:Transfer impedance solution} for details)
\begin{equation}
    Z_{21}(\omega) = iZ_0 \frac{\sin \! \left( \frac{\omega \ell_c}{v} \right) \cos \left( \frac{\pi \omega}{2\omega_n}\right)} {\cos \! \left( \frac{\pi\omega}{2\omega_p} \right) \cos \! \left(\frac{\pi\omega}{2\omega_r} \right)} \left(\frac{c_m}{c} \right) \textrm{,}
\label{eq:Z21 for homogeneous coupling}
\end{equation}
where $c=1/(Z_0 v)$ is the capacitance-to-ground per length of the CPW lines. The explicit dependence on the coupling inductance per length $l_m$ has dropped out because in a homogeneous medium it is fully determined by the other line properties~\cite{paul2007analysis}. Here, it satisfies $l_m=Z_0^2c_m$. The transfer impedance solution has its first zero at the frequency $\omega_n$, which takes the form
\begin{equation}
\omega_n = \frac{\pi v}{2\left(\ell_r^s + \ell_c + \ell_p^s\right)} \textrm{.}
\label{eq1:notch_frequency}
\end{equation}
\indent This frequency matches the anti-resonance of the effective $\lambda/2$ resonator with length $\ell_r^s + \ell_c + \ell_p^s$ that is formed between the shorted ends of the readout and filter resonators as depicted in Fig.~\ref{fig:Useful coupled resonator diagrams}(b). Due to its independence from the lengths $\ell_r^o$ and $\ell_p^o$, the frequency $\omega_n$ can be tuned independently from the readout- and filter-resonator frequencies and can be designed at both positive and negative detunings from these modes. As a result, it is well suited for use as a notch filter to suppress the Purcell relaxation channel for a qubit in both positive and negative detuning regimes. 
\begin{figure*}
\includegraphics[width=0.985\textwidth,height=0.49\textheight,keepaspectratio]{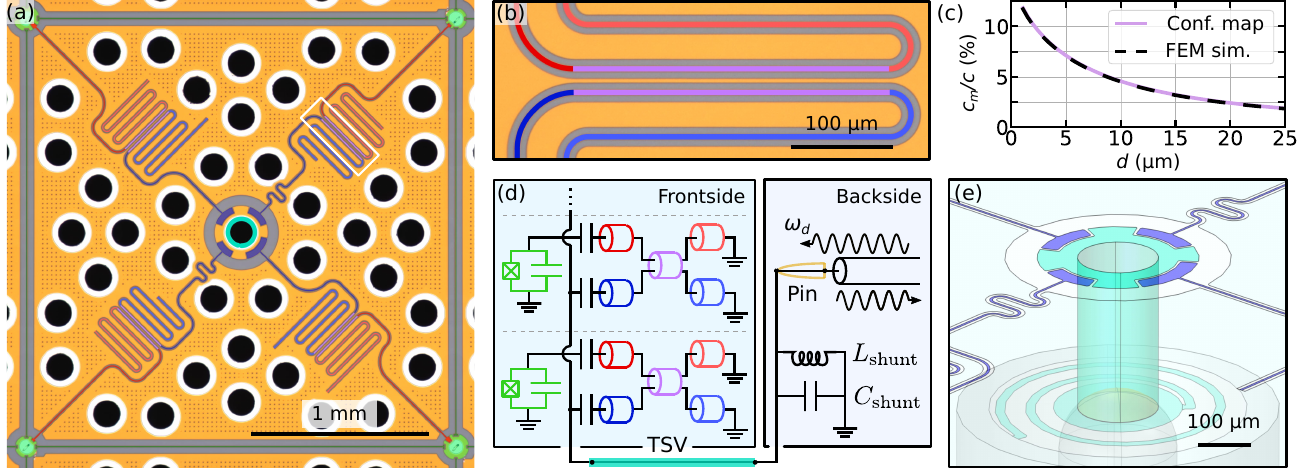}
\caption{\label{fig:Device plot} (a)~False-colored image of a four-qubit unit cell from a 16-qubit device. The four transmon qubits~(green) are coupled to their nearest neighbors by capacitive buses~(dark green). The four $\lambda/4$ readout resonators~(red) couple to $\lambda/4$ dedicated filter resonators~(blue), which couple to a shared readout line on the circuit backside through a TSV~(teal). (b)~Magnification of the region enclosed by the rectangle in~(a), showing the MTL section (purple) that couples the readout resonator and filter resonator. The width of the ground strip separating the two center lines is $d=\SI{3.8}{\micro\meter}$. (c)~Ratio of the per-length capacitances $c_m$ and $c$ as a function of the ground strip width $d$, determined numerically using a conformal-mapping technique~(solid) and from finite-element~(FEM) simulation using COMSOL~\cite{COMSOL}~(dashed). (d)~Circuit diagram of the readout configuration for the unit cell. (e)~Schematic three-dimensional transparent image of the TSV structure~(teal) that routes signals from a pogo pin~(yellow) on the circuit backside to the filter resonators on the circuit frontside. The TSV structure features a spiral inductor patterned on the circuit backside, which screens the parasitic shunt capacitance of the TSV structure.}
\end{figure*}
\\
\indent In order to simplify the analysis of the coupled resonators and to determine an expression for the coupling strength between them, we find an approximate mapping from the distributed-circuit representations in Fig.~\ref{fig:equivalent circuits}(a) to the lumped-element representations in Fig.~\ref{fig:equivalent circuits}(b). These lumped-element circuits accurately reproduce the electrical responses of the distributed circuits at frequencies around the $\lambda/4$ modes. The equations governing the mapping are provided in Appendix~\ref{appendix:Deriving equivalent circuits and Interaction Hamiltonian}. The $\lambda/4$ modes are converted to parallel LC resonators, the lumped-element capacitance $C_J$ in Fig.~\ref{fig:equivalent circuits}(a)(i) is transformed to the capacitance $\tilde{C}_J$, and the MTL coupler in Fig.~\ref{fig:equivalent circuits}(a)(ii) is transformed to a parallel LC resonator with capacitance $C_n$ and inductance $L_n$, satisfying $1/\sqrt{L_n C_n}=\omega_n$. 
\\
\indent The $Z_{21}(\omega)$ values for the equivalent lumped-element circuits and the exact solution to the distributed circuits are compared in Fig.~\ref{fig:equivalent circuits}(c) for a given set of parameters, showing excellent correspondence at frequencies around the notch frequency $\omega_n$ and the $\lambda/4$ modes. We infer that the low-energy Hamiltonian for the distributed circuit closely matches the  equivalent circuit Hamiltonian given the close correspondence of their electrical responses. The interaction Hamiltonian for the equivalent circuit with the MTL coupler depicted in Fig.~\ref{fig:equivalent circuits}(b)(ii) takes the form (see Appendix~\ref{appendix:Deriving equivalent circuits and Interaction Hamiltonian} for the derivation)
\begin{align}
& \hatH_{\mathrm{int}} = \hbar J (\hat{a}_r^\dag \hat{a}_p + \hat{a}_r \hat{a}_p^\dag) \textrm{,} \\
& J = \overline{\omega}_{rp}\frac{\pi^2}{32} \frac{ \left(\frac{\overline{\omega}_{rp}}{\omega_n} - \frac{\omega_n}{\overline{\omega}_{rp}} \right)^{\! 3}}{\cos^2 \! \left(\frac{\pi\omega_n}{2\overline{\omega}_{rp}} \right)} \left(\frac{c_m}{c} \right) 
\sin \! \left(\frac{\omega_n\ell_c}{v} \right) \textrm{,}
    \label{eq2:J_coupling}
\end{align}
where $\overline{\omega}_{rp} = (\omega_r + \omega_p)/2$ is the average of the readout- and filter-resonator frequencies, and we have taken $|\omega_r - \omega_p| \ll \omega_n$. The terms $\hat{a}_{r(p)}^\dag$ and $\hat{a}_{r(p)}$ are the creation and annihilation operators for the readout- and filter-resonator modes, and we have used the rotating-wave approximation (RWA) to drop the non-photon-number-preserving terms. The exchange coupling $J$ is suppressed as the notch frequency approaches $\overline{\omega}_{rp}$, scales linearly with the capacitance ratio $c_m/c$, and similarly scales linearly with the coupled-section length $\ell_c$, given that $\omega_n  \ell_c \ll v$. The capacitance ratio $c_m/c$ can be determined efficiently using a conformal-mapping technique~\cite{ghione1994efficient}, which allows $J$ to be predicted quickly for a given set of line dimensions. The dependence of $c_m/c$ on the width $d$ of the ground strip separating the coupled lines is plotted in Fig.~\ref{fig:Device plot}(c). The notch frequency is to a good approximation independent of $d$, per Eq.~\eqref{eq1:notch_frequency}. Thus, this width provides a parameter to tune the $J$ coupling independently of the notch frequency.
\subsection{Purcell-filtering enhancement}
In order to quantify the effect of the notch on the Purcell filtering, we compare the predicted Purcell-limited energy relaxation times for a qubit that is coupled to the readout line through a readout resonator and filter resonator that are MTL coupled and capacitively coupled. The coupling strength $J$ is assumed to be the same. We define the Purcell-filtering enhancement factor $\xi$ as the ratio of these relaxation times. Employing the equivalent circuits from the previous section, the enhancement factor is given by~(see Appendix~\ref{appendix:Purcell filtering enhancement expressions} for the derivation)
\begin{equation}
     \xi = \frac{1}{4} \frac{\omega_q^2}{\Delta_{qn}^2}\left(1-\frac{\omega_n^2}{\overline{\omega}_{rp}^2} \right) ^{\!\! 2} \textrm{,}
    \label{eq:Notch filter enhancement factor}
\end{equation}
where $\omega_q$ and $\Delta_{qn} = \omega_q - \omega_n$~(${\Delta_{qn} \ll \omega_n}$) are the qubit frequency and detuning from the notch, respectively. The enhancement diverges like $1/\Delta_{qn}^2$ as the qubit approaches the notch frequency. The expression can also be rearranged into a formula for the bandwidth $B$ around the notch where the enhancement to the Purcell-limited relaxation time of the qubit is at least a factor of $\xi$,
\begin{equation}
     B \approx \frac{\omega_n}{\sqrt{\xi}}\left|1-\frac{\omega_n^2}{\overline{\omega}_{rp}^2} \right| \textrm{.}
    \label{eq:Notch filter bandwidth}
\end{equation}
This expression predicts that, for the device parameters in the next section, the Purcell filtering is enhanced by at least two orders of magnitude over bandwidths exceeding \SI{200}{\mega\hertz} around the notch frequency.

%% file: main_body/Section3.tex
\section{Experimental Results}
\begin{figure}
\includegraphics[width=0.475\textwidth,height=0.49\textheight,keepaspectratio]{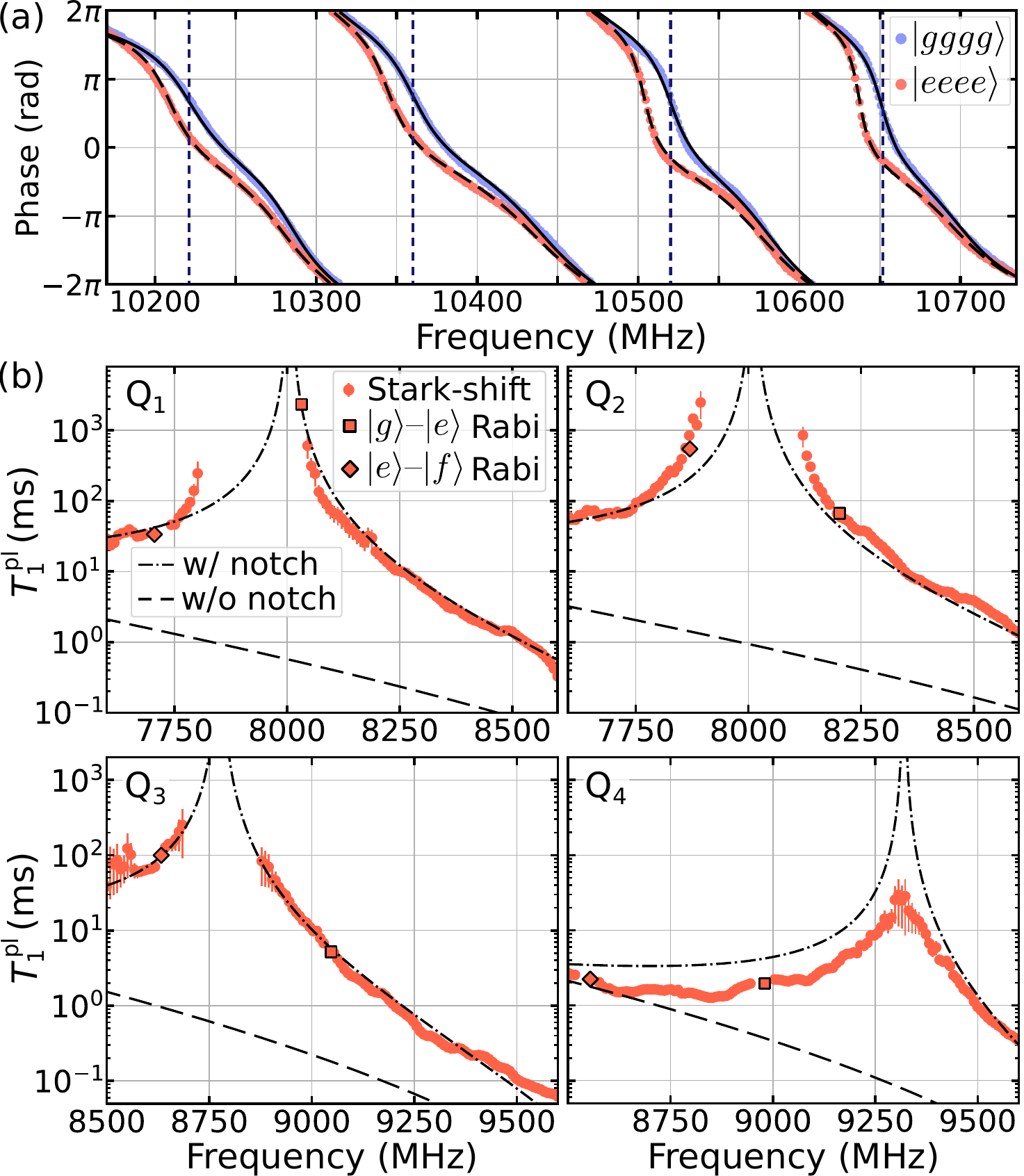}
\caption{\label{fig:fitted_spectroscopy_experiments} (a)~Reflection spectra of the four readout-resonator and filter-resonator pairs for the four qubits prepared in the $|g\rangle$~(blue) and $|e\rangle$~(red) states. The phase is wrapped modulo $4\pi$ to better visualize the response of each readout- and filter-resonator pair. The solid and dashed lines show the fitted phase response for the $|g\rangle$- and $|e\rangle$-state preparations, respectively. The vertical dashed lines show the four readout mode frequencies $\omega_{\textrm{ro}}^g$. (b)~Purcell-limited relaxation time $T_1^\mathrm{pl}$ of the four qubits as a function of frequency inferred from the Rabi oscillations driven at the qubit $|g\rangle$--$|e\rangle$ and $|e\rangle$--$|f\rangle$ transition frequencies~(squares and diamonds, respectively) and the ac Stark shift induced by the drive detuned from these transitions~(dots). The dot-dashed and dashed black lines show the prediction of a circuit model with and without the intrinsic notch filtering, respectively.}
\end{figure}
\subsection{Purcell protection of the qubit relaxation}
\begin{figure}
\includegraphics[width=0.48\textwidth,height=0.49\textheight,keepaspectratio]{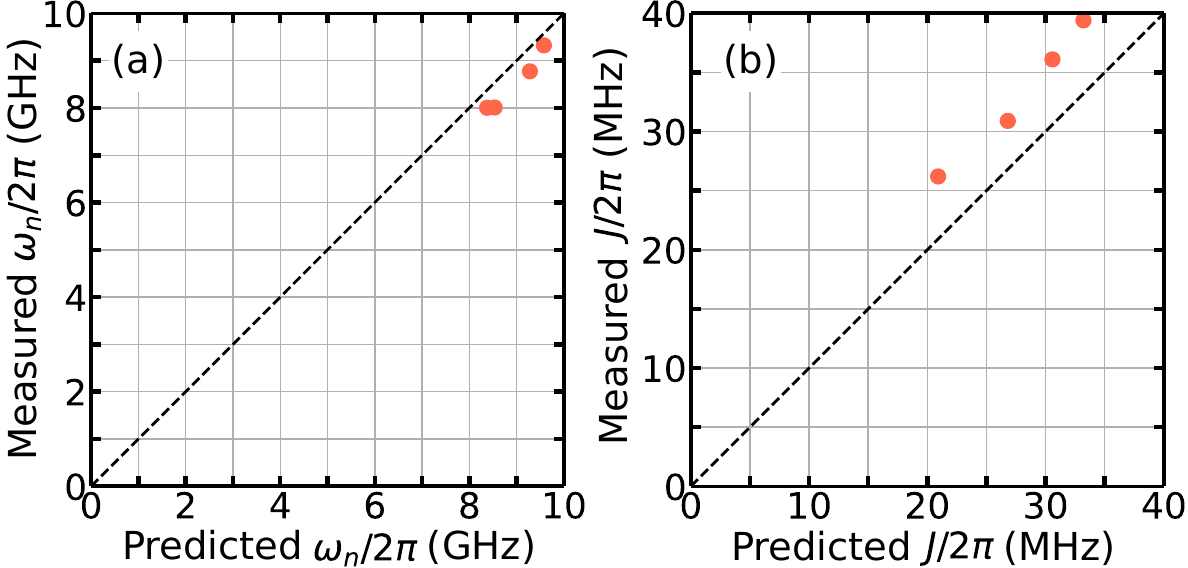}
\caption{\label{fig:Experiment_vs_prediction}Evaluation of the symbolic solutions for (a)~the predicted notch frequency $\omega_n$ and (b) the coupling strength $J$. The measured values are compared to the predictions of Eqs.~\eqref{eq1:notch_frequency} and~\eqref{eq2:J_coupling} for the four readout- and filter-resonator pairs.}
\end{figure}

We fabricated a 16-qubit device with each qubit coupled to a readout resonator and dedicated filter resonator featuring MTL coupling and notch frequencies that were tuned to the respective target qubit frequencies. The device architecture is based on a four-qubit unit cell that is tiled to create an array of fixed-frequency transmon qubits with nearest-neighbor couplings. Figure~\ref{fig:Device plot}(a) shows one of the four-qubit unit cells, featuring TiN-based transmon qubits and resonators, and superconducting through-silicon vias~(TSVs) metalized with Al. The TSVs connect the frontside ground planes via a common backside ground plane, and as a result airbridges~\cite{chen2014fabrication} are not used. Readout is performed by a reflection measurement through a pogo pin which contacts a TSV on the backside of the chip. Signals are routed to the filter resonators through the TSV as depicted in Fig.~\ref{fig:Device plot}(d). Figure~\ref{fig:Device plot}(e) shows a scale diagram of the through-substrate readout design. The TSV structure has a large ($\sim\SI{230}{\femto\farad}$) parasitic shunt capacitance to ground, which reduces the external linewidth of the filter resonators. In order to mitigate this, a spiral shunt inductor is patterned on the circuit backside to screen the shunt capacitance around the readout frequencies (see Appendix~\ref{appendix:Semi-classical model for the readout system dynamics} for details).
\\
\indent Figure~\ref{fig:fitted_spectroscopy_experiments}(a) shows the reflected phase response of a signal applied to the readout pogo pin on the circuit backside. Due to the relatively small detuning between the filter resonators compared to their external linewidths, the parameters of the four readout resonators and four filter resonators were fitted simultaneously to a reflection model that encompassed all eight resonators in the four-qubit unit cell (see Appendix~\ref{appendix:Semi-classical model for the readout system dynamics} for details). By fitting the $|g\rangle$- and $|e\rangle$-state responses simultaneously, the dispersive shifts of the readout resonators were also determined. The readout-mode parameters determined from the fit, together with the qubit parameters for qubits Q$_1$ to Q$_4$ determined from their spectroscopy, are presented in Table~\ref{tab:Readout parameters}. Here, the readout modes refer to the hybridized modes formed between the readout and filter resonators which were addressed during the qubit readout (see Appendix~\ref{appendix:Semi-classical model for the readout system dynamics} for details). The bare readout- and filter-resonator parameters are also presented in Table~\ref{tab:Bare readout parameters - appendix} in Appendix~\ref{appendix:Semi-classical model for the readout system dynamics}.
\begin{table}
\centering
\caption{\label{tab:Readout parameters}Qubit and readout parameters. Here, $\omega_q$ and $\alpha$ are the qubit frequency and anharmonicity, respectively. The terms $\omega_{\textrm{ro}}^g$ and $\omega_{\textrm{ro}}^e$ are the hybridized readout mode frequencies given the qubit is in the $|g\rangle$ and $|e\rangle$ state, respectively, and $\chi_{\textrm{ro}}$ is the readout mode dispersive shift defined $\omega_{\textrm{ro}}^e \equiv \omega_{\textrm{ro}}^g+2\chi_{\textrm{ro}}$. The term $T_{1}^\mathrm{pl}$ is the Purcell-limited relaxation time evaluated at the qubit frequency and $\xi$ is the predicted Purcell-filtering enhancement factor due to the intrinsic notch filter.}
\begin{ruledtabular}
\begin{tabular}{cccccccc}
& $\omega_q/2\pi$ & $\alpha/2\pi$ & $\omega_{\textrm{ro}}^g/2\pi$ & $\kappa_{\textrm{ro}}^g/2\pi$ &  $\chi_{\textrm{ro}}/2\pi$ & $T_1^\mathrm{pl}$ & $\xi$\\ 
 & (\si{\mega\hertz}) & (\si{\mega\hertz}) &  (\si{\mega\hertz}) & (\si{\mega\hertz}) & (\si{\mega\hertz}) & (ms) & \\ \hline
Q$_1$ & $8032$ & $-326$ &  $10221$ & $42$ & $-5.9$ & $2300$ & $\sim$3000 \\ 
Q$_2$ & $8189$ & $-333$ &  $10360$ & $34$ & $-7.8$ & $70$ & 85\\ 
Q$_3$ & $9046$ & $-414$ & $10520$ & $24$ & $-8.3$ & $5$ & 25\\  
Q$_4$ & $8980$ & $-425$ & $10652$ & $19$ & $-7.1$ & $2$ & 10 \\ 
\end{tabular}
\end{ruledtabular}
\end{table}
\\
\indent The Purcell protection of each qubit was characterized following the method in Ref.~\citenum{sunada2022fast}. First, the drive power $P$ incident on the device was calibrated by measuring the ac Stark shift of the qubit caused by a readout tone (see Appendix~\ref{appendix:Ac Stark shift spectroscopy measurement} for details). The drive amplitude $\Omega$ incident on the qubit was then determined by two methods: (i)~by measuring the induced Rabi-oscillation frequency in the qubit when the drive was on-resonance with the $|g\rangle$--$|e\rangle$ and $|e\rangle$--$|f\rangle$ transitions and (ii)~by measuring the drive-induced Stark shift on the qubit when the drive was detuned from these transitions. The Purcell-limited relaxation time $T_1^{\textrm{pl}}$ over a range of frequencies was then determined from the ratio of $P$ and $\Omega^2$ (see Appendix~\ref{appendix:Ac Stark shift spectroscopy measurement} for details). The Purcell-limited relaxation times for the four qubits are shown in Fig.~\ref{fig:fitted_spectroscopy_experiments}(b). The predicted values using an equivalent-circuit model of the readout with and without the intrinsic notch filtering are overlaid (see Appendix~\ref{appendix:Purcell filtering enhancement expressions} for details). The notch frequency is the only fitting parameter in these models, with the circuit otherwise determined from the qubit and readout parameters. The Purcell-limited relaxation times at the $|g\rangle$--$|e\rangle$ transition frequency, and the predicted relaxation enhancement factors $\xi$ are presented in Table~\ref{tab:Readout parameters}. The evaluated Purcell-limited relaxation times all exceed $\SI{2}{\milli\second}$ and are predicted to be enhanced by at least an order of magnitude due to the intrinsic notch filtering. In the case of qubit Q$_1$, the notch frequency is nearly resonant with the qubit frequency, and as a result the measured value of $T_1^{\textrm{pl}}$ approaches $\SI{2.5}{\second}$ even though the effective readout linewidth is $\SI{42}{\mega\hertz}$, demonstrating the effective elimination of the Purcell relaxation channel.
\\
\indent Figure~\ref{fig:Experiment_vs_prediction} compares the measured notch frequencies and coupling strengths $J$ to the predictions of Eqs.~\eqref{eq1:notch_frequency} and~\eqref{eq2:J_coupling}, respectively. The phase velocity $v$ and the capacitance ratio $c_m/c$ in the symbolic expressions were determined numerically by conformal mapping. The measured values of the notch frequencies $\omega_n$ were used in Eq.~\eqref{eq2:J_coupling} to calculate the $J$ values, in order to assess the accuracy of the two expressions independently. The measured notch frequencies and $J$ couplings on average differ from the predicted values by 5\% and 19\%, respectively. The underestimation of the coupling strengths is in part due to the assumption that coupling only occurs along the MTL-section-length $\ell_c$ as defined in Fig.~\ref{fig:Device plot}(b), since this neglects the coupling between the curved sections of the lines.

%% file: main_body/Section4.tex
\subsection{Multiplexed fast readout}
\begin{figure}
\includegraphics[width=0.48\textwidth,height=0.49\textheight,keepaspectratio]{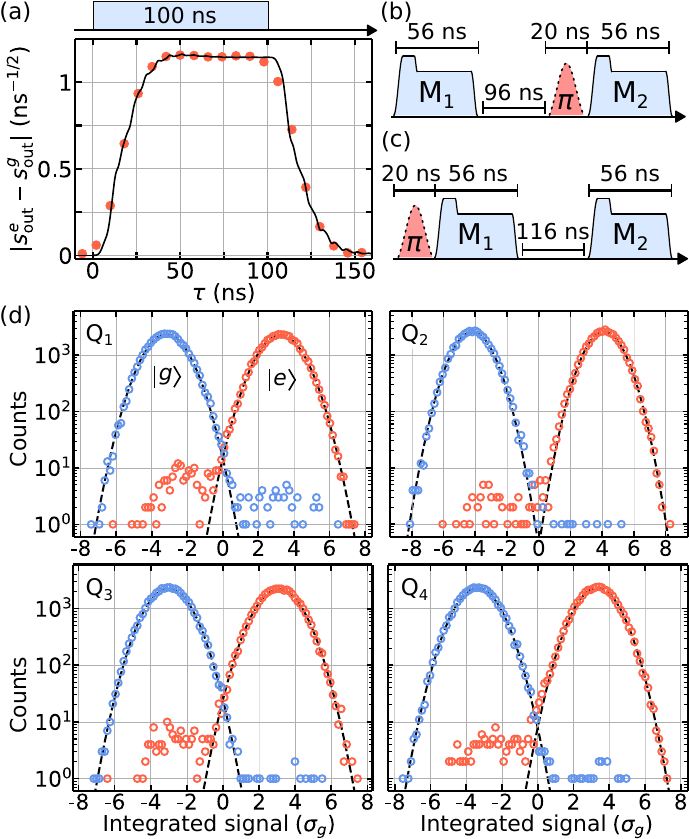}
\caption{\label{fig:fast_readout_experiment} Demonstration of fast readout. (a)~Time dependence of the output-field separation $S(t)\equiv|s_\mathrm{out}^e(t)-s_\mathrm{out}^g(t)|$ in response to a 100-ns-long rectangular readout pulse at the optimal readout frequency of qubit Q$_2$, averaged over $4\times10^{4}$ repetitions. The solid black line is the prediction of the semi-classical model for the coherent output fields. (b)~and (c)~Readout pulse sequences used to determine the assignment and QND fidelities, respectively. The readout sequences are performed with and without the $20$-ns-long $\pi$-pulse. (d)~Histograms of simultaneous time-integrated single-shot signals for the four qubits projected along the principal axes in the in-phase and quadrature~(IQ) plane. Dashed lines show the fits of the $|g\rangle$- and $|e\rangle$-state signal distributions to Gaussian distributions. The horizontal axes are normalized by the standard deviation $\sigma_g$ of the $|g\rangle$-state Gaussian distribution.
}
\end{figure}
Figure~\ref{fig:fast_readout_experiment}(a) shows the time dependence of the separation $S(t)\equiv|s_\mathrm{out}^e(t)-s_\mathrm{out}^g(t)|$ of the qubit-state-dependent coherent output fields $s_\mathrm{out}^{g(e)}$ from the device in response to a 100-ns-long rectangular tone applied at the optimal readout frequency of qubit Q$_2$. The measured state-dependent heterodyne signals were converted to the separation through the relationship $\Gamma_m=S_{\mathrm{ss}}^2/2$ between the measurement-induced dephasing rate $\Gamma_m$ and the steady-state separation $S_{\mathrm{ss}}$~\cite{Gambetta2006}. The measurement displays a short transient period of around \SI{50}{\nano\second} before the output field carries the maximum information about the qubit state. A semi-classical model for the output-field separation (see Appendix~\ref{appendix:Semi-classical model for the readout system dynamics}) evaluated using the independently determined readout- and filter-resonator parameters shows good correspondence to the measured dynamics.
\\
\indent The rapid rise in output-field separation enabled the tune-up of a $56$-\si{\nano\second} multiplexed readout on the four qubits with high fidelity. For single-shot readout, a flux-driven impedance-matched parametric amplifier~\cite{mutus2014strong} was used in the phase-preserving mode of operation. At the optimal point, the quantum efficiency $\eta$ of the readout chain was determined to be $\{27, 46, 40, 42\}$\si{\percent} for qubits Q$_1$ to Q$_4$, respectively, by comparing the measurement-induced dephasing to the measured signal-to-noise ratio (SNR)~\cite{bultink2018general}.
\\
\indent Figures~\ref{fig:fast_readout_experiment}(b) and~\ref{fig:fast_readout_experiment}(c) show the pulse sequences used to benchmark the readout assignment fidelity $F$ and the quantum non-demolition (QND) fidelity $F_Q$, respectively. The pulse sequences were carried out on the four qubits simultaneously and repeated $4\times10^4$ times both with and without the $\pi$-pulse. The two-step measurement pulses featured an initial 14-\si{\nano\second} flat top with a 6-\si{\nano\second} raised-cosine edge and a flat-top amplitude between 1.35--1.4 times higher than that of the subsequent flat plateau in order to increase the output-field separation at early times. The steady-state photon-number populations of the readout resonators during the main pulse step, $n_{r,j}^g$ (given the qubit is in the $|g\rangle$ state), are provided in Table~\ref{tab:readout benchmarking} as a fraction of the critical photon number~\cite{Blais2020Circuit}.  The acquired measurement signals were averaged over an optimally-weighted integration time window of $\SI{56}{\nano\second}$. The qubit states were assigned from the integrated signals using a two-state logistic-regression discrimination model~\cite{pedregosa2011scikit} that was trained on a prior dataset of $2\times10^4$ single-shot measurements. The assignment fidelities were determined using the definition $F\equiv \left[P_{0}\left(g_2|g_1\right)+P_{\pi}\left(e_2|g_1\right)\right]/2$, where $P(a|b)$ denotes the probability of event $a$ given event $b$. The subscripts~$1$ and~$2$ refer to the first and second measurement outcomes, and the subscripts $\pi$ and $0$ refer to the pulse sequences in Fig.~7(b) with and without the $\pi$-pulse. The QND fidelities were determined from the QND pulse sequences in Fig.~7(c) using the definition $F_{\mathrm{Q}}\equiv \left[P_{0}\left(g_2|g_1\right)+P_{\pi}\left(e_2|e_1\right)\right]/2$. The resulting assignment and QND errors are summarized in Table~\ref{tab:readout benchmarking} along with the predicted coherence-limited error imposed by qubit relaxation.
\\
\indent Figure~\ref{fig:fast_readout_experiment}(d) shows the histograms formed by projecting the single-shot IQ points from the second measurement of the assignment fidelity pulse sequence along the principal axis. The results are post-selected to ensure the qubit was in the $|g\rangle$ state prior to the $\pi$-pulse. The histogrammed signals closely match Gaussian distributions with outliers that are discussed below. From the overlap of the Gaussian fits the state-separation errors $\varepsilon_{\textrm{sep}}$ are calculated and summarized in Table~\ref{tab:readout benchmarking}. 
\\
\indent The average assignment fidelity for the $56$-\si{\nano\second} simultaneous measurement is $99.77\%$, representing a significant advancement for fast multiplexed qubit readout~\cite{Heinsoo2018}. Focusing on qubit Q$_2$, the separation error was negligible and the assignment fidelity of $99.91(1)\%$ reached the coherence limit. The probability of measurement-induced excitation from $|g\rangle$ to $|e\rangle$ was low with $P_{0}\left(e_2|g_1\right)=0.03\%$. In Appendix~\ref{appendix: Readout analysis}, the IQ plot is examined, showing that less than $0.01\%$ of measurements exhibited discernible excitation to leakage states~\cite{sank2016measurement, shillito2022dynamics, dumas2024unified}. The QND fidelity was $0.37\%$ below the coherence limit, implying some activation of measurement-induced leakage and measurement-induced relaxation~\cite{harrington2017quantum, thorbeck2024readout}. On the other hand, the assignment and QND fidelities for qubit Q$_1$ (which had the lowest quantum efficiency) and Q$_3$ were markedly lower than their respective coherence limits, implying that the readout pulse caused significant measurement-induced excitation and relaxation. The IQ plot for Q$_1$ reveals measurement-induced excitation to leakage states that contributed significantly to the assignment error (see Appendix~\ref{appendix: Readout analysis}). This could be countered with improved parametric amplification at the readout frequency of Q$_1$, which would facilitate readout at a reduced photon number. Pseudo-syndrome detection benchmarking could be performed to build a more complete picture of the measurement-induced leakage~\cite{Hazra2024benchmarking}. Additional optimizations could be achieved by more advanced readout pulse shaping~\cite{gautier2024optimal, jerger2024dispersive}. Further increasing the assignment fidelities beyond $99.9\%$ will require increasing the qubit relaxation times. The averaged qubit relaxation times $T_1$ of qubit Q$_1$ to Q$_4$ in the current experiments were $\{45(6), 26(1), 38(1), 34(1)\}~\si{\micro\second}$ at the time of the readout benchmarking.
\begin{table}
\centering
\caption{\label{tab:readout benchmarking}Readout results. Here, $n_{r,j}^g/n_{\textrm{crit}}$ represents the ratio of the steady-state photon number in the readout resonator during the readout pulse to the critical photon number. The terms $\varepsilon$ and $\varepsilon_\mathrm{Q}$ are the assignment and QND errors, respectively, defined $\varepsilon \equiv 1-F$ and $\varepsilon_Q \equiv 1-F_Q$. The terms $\varepsilon_\mathrm{cl}$ and $\varepsilon_\mathrm{cl}^Q$ are the coherence-limited error budgets on the assignment and QND fidelity, estimated from measurements of the qubit relaxation time carried out within an hour of the readout benchmarking. Details of the coherence limits and separation error calculations are given in Appendix~\ref{appendix: Readout analysis}} 
\begin{ruledtabular}
\begin{tabular}{ccccccc}
& $n_{r,j}^g/n_{\textrm{crit}}$ & $\varepsilon$ (\si{\percent}) & $\varepsilon_\mathrm{cl}$ (\si{\percent}) & $\varepsilon_Q$ (\si{\percent}) & $\varepsilon_\mathrm{cl}^Q$ (\si{\percent}) & $\varepsilon_{\textrm{sep}}$ (\si{\percent}) \\ 
\hline
Q$_1$ & $1.50$ & $0.35(2)$ & 0.06 & $1.91(5)$ & $0.19$ & 0.08 \\
Q$_2$ & $1.05$ & $0.09(1)$ & 0.11 & $0.70(3)$ & $0.33$  & $<$$0.01$ \\
Q$_3$ & $0.82$ & $0.29(2)$ & 0.07 &  $1.10(4)$ & $0.23$ & 0.13 \\
Q$_4$ & $0.89$ & $0.20(2)$ & 0.08 & $0.55(3)$ & $0.25$ & 0.04 \\
\end{tabular}
\end{ruledtabular}\end{table}
\\
\indent The results demonstrate that tailoring readout modes to have large external linewidths and dispersive shifts is an effective strategy to increase assignment fidelities, in agreement with Refs.~\citenum{Walter2017} and~\citenum{swiadek2023enhancing}. A potential drawback of this approach is that it increases the susceptibility of qubits to dephasing due to the noise-photon population in the readout resonator~\cite{bertet2005dephasing, yan2018distinguishing, wang2019cavity}. However, we expect some degree of protection from this dephasing channel in this device due to the high readout resonator frequencies, which were all above \SI{10}{\giga\hertz}. This is because a thermal photon population in the readout resonator is expected to follow Bose-Einstein statistics~\cite{goetz2017photon} and thus be exponentially suppressed with increasing frequency at a given temperature. The Hahn-echoed pure dephasing times $T_{2\mathrm{echo}}$ of qubits Q$_1$ to Q$_4$ were found to be $\{61,55,152,77\}~\si{\micro\second}$. Attributing this dephasing entirely to the noise-photon population $n_{r,j}^{\textrm{noise}}$ of the readout resonators sets upper bounds of ${n_{r,j}^{\textrm{noise}} \leq \{3.1,3.2,1.0,2.4\} \times 10^{-4}}$ (see Appendix~\ref{appendix:Semi-classical model for the readout system dynamics} for details). These values correspond to effective temperatures of between 55 to \SI{62}{\milli\kelvin} as a result of the high readout resonator frequencies. It may be possible to further suppress the sensitivity to noise-photon induced dephasing by introducing weak anharmonicity to the filter resonators~\cite{sunada2024photon}.

%% file: main_body/Section5.tex
\section{Conclusion}
We have shown that by coupling a quarter-wavelength readout resonator and filter resonator together, both capacitively and inductively, it is possible to engineer an  intrinsic notch filter that can provide strong auxiliary Purcell filtering to the qubit. An equivalent-circuit model has been introduced and used to find symbolic expressions for the notch filtering and for the coupling strength between the readout and filter resonators. We have experimentally verified these formulas in a device with four-to-one readout multiplexing. The notch filter broadens the range of linewidths for the readout mode that are possible without impacting the qubit relaxation time. By designing readout modes with external linewidths as large as $\SI{42}{\mega\hertz}$, we tuned up a fast (56-ns) multiplexed readout across four qubits and benchmarked a state-of-the-art average simultaneous assignment fidelity of $99.77\%$, with the highest qubit assignment fidelity equal to $99.91\%$. The compact coupled resonators with intrinsic notch filtering offer strong Purcell protection and are straightforward to design, making this an attractive circuit component for superconducting qubit readout.

%% file: Acknowledgments.tex
\begin{acknowledgments}
The work was supported in part by the Ministry of Education,
Culture, Sports, Science and Technology (MEXT) Quantum Leap Flagship Program~(Q-LEAP)~(Grant No.~JPMXS0118068682), and the JSPS Grant-in-Aid for Scientific Research~(KAKENHI)~(Grant No.~JP22H04937).
\end{acknowledgments}

%% file: appendix_body/Supplementary_A.tex
\section{Device and experimental setup}
\label{appendix:Device setup}
\begin{figure}
\includegraphics[width=0.475\textwidth,height=0.49\textheight,keepaspectratio]{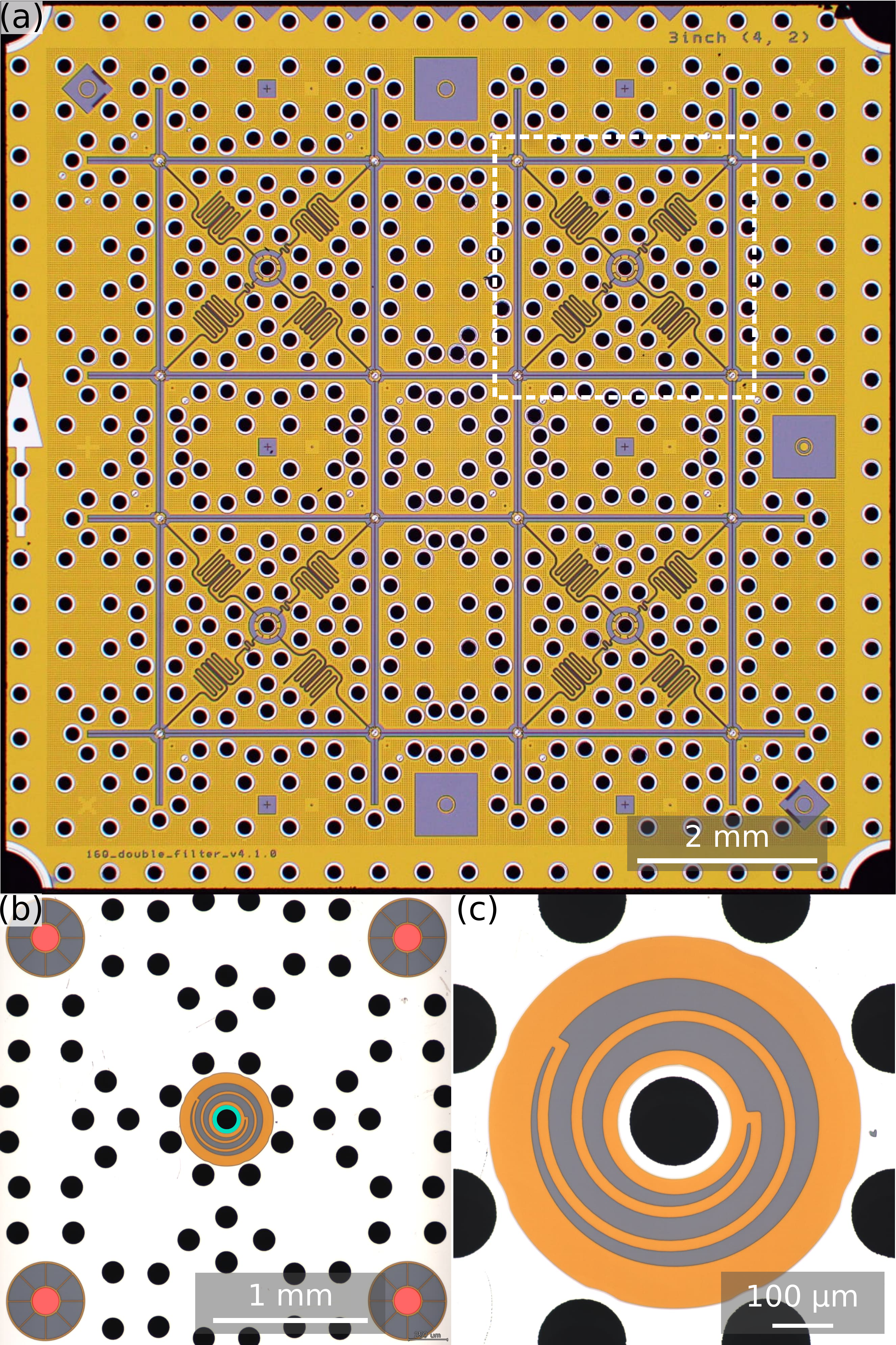}
\caption{\label{fig:16_q_device} Images of the 16-qubit device. (a) Photograph of the 16-qubit chip. The four-qubit unit cell measured in this work is highlighted. (b) False-color image of the backside of the four-qubit unit cell, showing the contact pads where pogo pins contact the chip for qubit control (red) and for multiplexed qubit readout (teal). (c) Image of the TSV contacted by the readout pogo pin, showing the TiN spiral shunt inductor.}
\end{figure}

\begin{figure}
\includegraphics[width=0.475\textwidth,height=0.49\textheight,keepaspectratio]{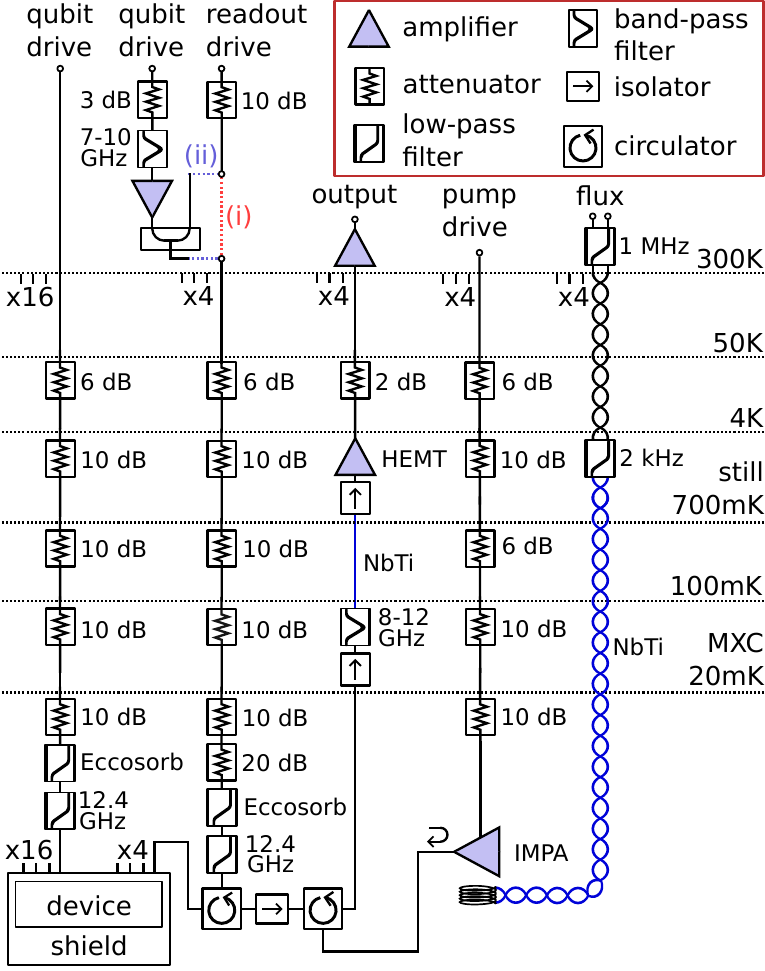}
\caption{\label{fig:fridge_diagram} Experimental setup. The readout drive line was wired in two configurations. In configuration~(i), the readout drive is connected directly to the input line. In configuration~(ii), it is combined with an amplified qubit drive before connecting to the input line. The IMPA amplifier refers to the impedance-matched parametric amplifier. The term MXC refers to the mixing chamber stage.}
\end{figure}

The four-qubit unit cell measured in this work is from the 16-qubit device shown in Fig.~\ref{fig:16_q_device}(a). The qubits are coupled to nearest neighbors (NN) by short transmission-line sections, resulting in an effective capacitive coupling between nearest neighbors. The average NN coupling strength is $\SI{7.0}{\mega\hertz}$, and the NN static-ZZ crosstalk ranges between $80$ to $\SI{200}{\kilo\hertz}$. The circuit backside where the qubit- and readout-drive wirings make contact with the chip is shown in Fig.~\ref{fig:16_q_device}(b). Figure~\ref{fig:16_q_device}(c) shows a close up of the TSV contacted by a readout pogo pin which routes readout signals to the circuit frontside.
\\
\indent
The device was cooled to $\sim$\SI{20}{\milli\kelvin} inside a Bluefors XLD1000 dilution refrigerator. The experimental setup is shown in Fig.~\ref{fig:fridge_diagram}. For the qubit and readout drives, the impedance-matched parametric amplifier pump drive, and the readout signal acquisition, a \mbox{QuEL-1} controller~(QuEL~Inc.~\cite{sumida2024qube1}) was used. For the experiments characterizing the Purcell filtering, the readout drive was combined with a qubit drive which was used to generate the Stark shifting tone. In this configuration, a room-temperature amplifier (\mbox{Mini-circuits ZVA-183W-S+}) was attached to the Stark-shifting qubit drive in order to increase the Stark shift induced in the qubit.

%% file: appendix_body/Supplementary_B.tex
\section{Transfer impedance between two MTL-coupled $\mathbf{\lambda/4}$ resonators}
\label{appendix:Transfer impedance solution}
\begin{figure}
\includegraphics[width=0.475\textwidth,height=0.49\textheight,keepaspectratio]{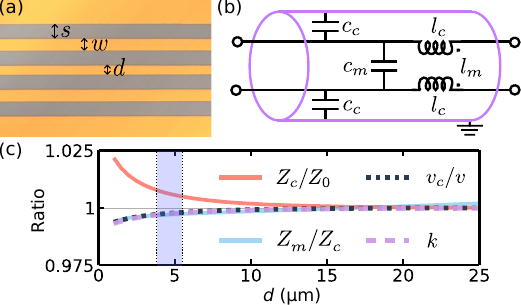}
\caption{\label{fig:weak coupling and homogeneous assumption} Analysis of coupled CPW lines. (a)~Image of the coupled CPW lines used to realize the MTL section. (b)~Definitions for the per-length parameters of the two-line MTL. (c)~Finite-element electrostatic simulation results using COMSOL for the coupled CPW lines showing dimensionless quantities of the MTL lines as a function of the width $d$ of the separating ground strip. The values of $d$ for the coupled CPW lines in this work lie inside the blue highlighted region.}
\end{figure}

\begin{figure}
\includegraphics[width=0.4\textwidth,height=0.4\textheight,keepaspectratio]{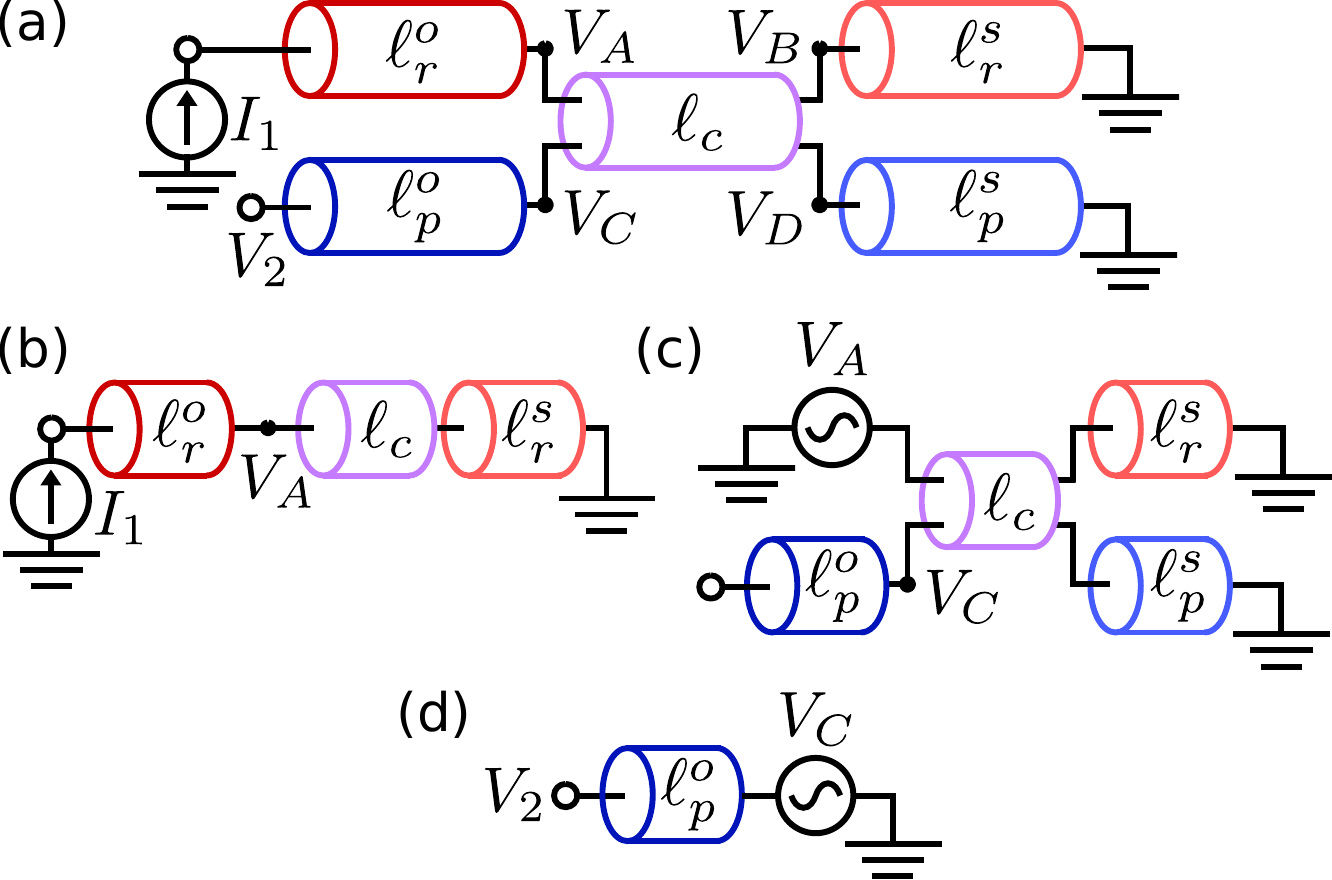}
\caption{\label{fig:Transfer impedance solving circuit} Solution steps for determining the transfer impedance $Z_{21}(\omega)$. (a)~Distributed-circuit model for two MTL-coupled $\lambda/4$ resonators. A current source drives port~1, and the goal is to determine the induced voltage $V_2$ at port~2. The induced voltages at the four ports of the MTL are defined as $V_A$ to $V_D$. (b)~Circuit used to find the voltage $V_A$ in terms of the current drive $I_1$. (c)~Circuit used to find the voltage $V_C$ in terms of the voltage drive $V_A$. (d)~Circuit used to find the voltage $V_2$ in terms of the voltage drive $V_C$.}
\end{figure}

\begin{figure}
\includegraphics[width=0.325\textwidth,height=0.325\textheight,keepaspectratio]{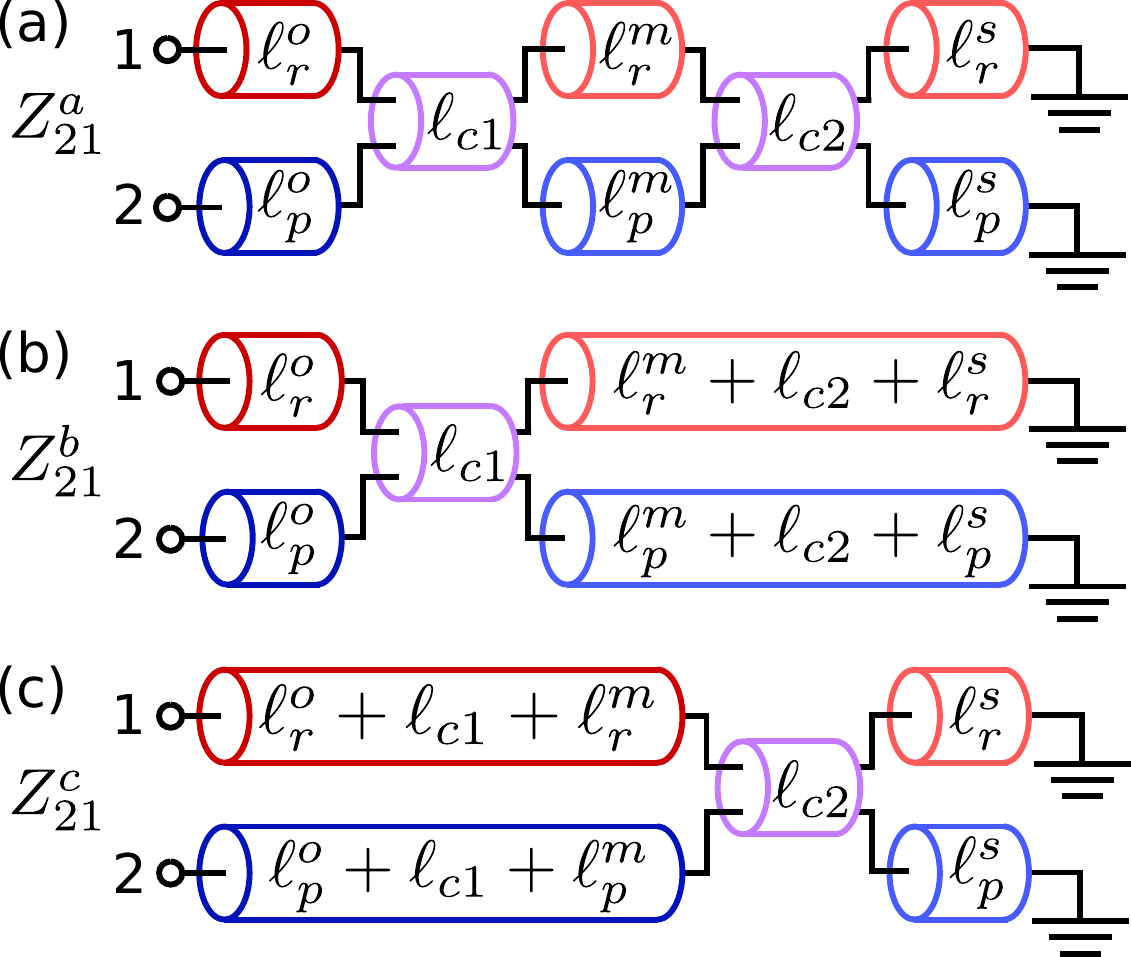}
\caption{\label{fig:transfer_impedance_multiple_MTLs} (a) Pair of $\lambda/4$ resonators coupled by two MTL sections. The transfer impedance is defined $Z_{21}^a$. (b)~and (c)~Same pair of resonators coupled by only the first and second MTL sections, respectively. Under the weak-coupling and consonant-line assumptions, the transfer impedances satisfy the relation $Z_{21}^a=Z_{21}^b+Z_{21}^c$.}
\end{figure}

In this section, we derive an analytic solution for the transfer impedance between two MTL-coupled $\lambda/4$ resonators. We outline the necessary assumptions for achieving a tractable solution and demonstrate that these assumptions are well justified for the coupled CPW lines in this work. We discuss the general solution as well as the specific cases of lumped-element capacitive coupling and MTL coupling in a homogeneous medium.
\\
\indent Figure~\ref{fig:weak coupling and homogeneous assumption}(a) shows an image of two coupled CPW lines of the kind used in this work to couple the readout and filter resonators. The key line dimensions are the ground separation $s=\SI{7.5}{\micro\meter}$, the center-line width $w=\SI{5}{\micro\meter}$, and the width of the ground strip that separates the two CPWs, $d$, which took values between $\SI{3.8}{\micro\meter}$ and $\SI{5.5}{\micro\meter}$.
\\
\indent The two coupled CPW lines are treated as an MTL with per-length parameters as defined in Fig.~\ref{fig:weak coupling and homogeneous assumption}(b). In order to find a tractable symbolic solution for the transfer impedance $Z_{21}(\omega)$ between two $\lambda/4$ resonators coupled together by this MTL, the following assumptions are applied:
{\allowdisplaybreaks
\begin{itemize} 
\item Cyclic-symmetric lines: The two lines of the MTL are both taken to have the same capacitance-to-ground and self-inductance per length, here denoted $c_c$ and $l_c$, respectively.
\item Weakly-coupled lines: An excitation applied to one line of the MTL will induce an excitation in the other line. The weak-coupling assumption amounts to neglecting the back-action of this induced excitation onto the first line and is valid when $l_m \ll l_c$ and $c_m \ll c_c$. In a homogeneous medium, this is relaxed to the less restrictive condition $k \equiv (1-\left(l_m/l_c\right)^2)^{1/2}\approx 1$~\cite{paul2002solution}.
\item Consonant lines: The two lines of the MTL have characteristic impedance and phase velocity defined as $Z_c \equiv \sqrt{l_c/(c_c+c_m)}$ and $v_c \equiv  1/\sqrt{l_c(c_c+c_m)}$, respectively. Here, the term consonant line is defined to mean that these quantities equal the characteristic impedance and phase velocity for the uncoupled sections of the $\lambda/4$ resonators, such that $Z_c = Z_0$ and $v_c = v$. 
\end{itemize}
}
\indent Figure~\ref{fig:weak coupling and homogeneous assumption}(c) presents COMSOL finite-element simulation results showing that the weak-coupling and consonant-line assumptions are well justified for the coupled CPW lines considered in this work. The simulation results also confirm the relationship $Z_m=Z_c$ for two coupled CPW lines, where $Z_m\equiv \sqrt{l_m/c_m}\,$. As discussed later in Appendix~\ref{appendix:Transfer impedance solution}2, this relationship is characteristic of lines that are embedded in a homogeneous medium.
\\ 
\indent Figure~\ref{fig:Transfer impedance solving circuit}(a) shows the distributed circuit used to model the coupled resonators, with a current source attached to port 1. The transfer impedance between ports~1 and~2 is defined as $Z_{21}(\omega)\equiv V_2(\omega)/I_1(\omega)\vert_{I_2=0}$. In order to solve for this, the problem is broken down into parts as depicted in Figs.~\ref{fig:Transfer impedance solving circuit}(b)--(d). First, the ratio $V_A(\omega)/I_1(\omega)$ is solved for, where $V_A(\omega)$ is the voltage at port A of the MTL. Using the weak-coupling assumption to neglect back-action and using the consonant-line assumption to set $Z_c=Z_0$ and $v_c=v$, the problem reduces to that of solving for the voltage at a distance $\ell_r^o$ from the open end of a shorted transmission line of total length $\ell_r=\ell_r^o+\ell_c+\ell_r^s$, as depicted in Fig.~\ref{fig:Transfer impedance solving circuit}(b). This yields
\begin{equation}
\frac{V_A(\omega)}{I_{1}(\omega)}= iZ_0\left[\tan \! \left(\frac{\omega \ell_r }{v} \right) \cos \! \left(\frac{\omega \ell_{r}^o}{v} \right) - \sin \! \left(\frac{\omega \ell_{r}^o}{v}  \right) \right]  \textrm{.} 
\label{eq:voltage to source current relation}
\end{equation}
\indent Next, the voltage $V_C(\omega)$ at port C of the MTL line is solved for in terms of the voltage $V_A(\omega)$, as depicted in Fig.~\ref{fig:Transfer impedance solving circuit}(c). Here, 
we make use of the general solution presented in Eq.~(26a) of Ref.~\citenum{paul2002solution} for a weakly-coupled two-line MTL terminated with arbitrary loads. We use the consonant-line assumption to set $Z_c=Z_0$ and $v_c=v$ and after some algebraic manipulations arrive at the following expression,
\begin{equation}
\frac{V_C(\omega)}{V_A(\omega)} = Z_0 \omega  \ell_c c_m  \frac{\left(A_{+} - A_{-}\right)\cos \! \left( \frac{\omega\ell_{p}^o}{v}\right)}{2\cos \! \left(\frac{\pi\omega}{2\omega_p}\right)\sin \! \left[\frac{\omega\left(\ell_r^s+ \ell_c\right)
}{v}\right]} \textrm{,}
\label{eq:voltage ratio solution under weak coupling assumption and consonant lines assumption}
\end{equation}
where
\begin{align}
& A_+ =  \left(1 + Z_m^2/Z_0^2\right) \sinc \! {\left(\frac{\omega \ell_c}{v}\right)}\cos \! \left[\frac{\omega\left(\ell_r^s+\ell_p^s+\ell_c\right)}{v}\right] \textrm{,}\\
& A_- = \left(1 - Z_m^2/Z_0^2\right) \cos \! \left[\frac{\omega \left(\ell_r^s - \ell_p^s \right)}{v} \right] \textrm{.}
\end{align}
Here, we restate that $Z_m \equiv \sqrt{l_m/c_m}\,$. Finally, the voltage $V_2$ is solved for in terms of the voltage $V_C$ as depicted in Fig.~\ref{fig:Transfer impedance solving circuit}(d), yielding
\begin{equation}
    \frac{V_2(\omega)}{V_C(\omega)} = \sec \! \left( \frac{\omega \ell_p^o}{v} \right)  \textrm{.} \label{eq:open_transmission_line_voltage_ratio}
\end{equation}
\indent The transfer impedance is found by combining these solutions:
\begin{equation}
    Z_{21}(\omega) = \left(\frac{V_2}{V_C}\right)\left(\frac{V_C}{V_A}\right)\left(\frac{V_A}{I_1}\right) \textrm{.}
    \label{eq:transfer impedance combination}
\end{equation}
\indent Substituting in Eqs.~\eqref{eq:voltage to source current relation},~\eqref{eq:voltage ratio solution under weak coupling assumption and consonant lines assumption} and~\eqref{eq:open_transmission_line_voltage_ratio}, the overall expression simplifies to yield the final result
\begin{equation}
    Z_{21}(\omega) = iZ_0^2\omega\ell_c c_m\frac{A_{+} - A_{-}}{2\cos \! \left(\frac{\pi\omega}{2\omega_r} \right) \cos \! \left(\frac{\pi\omega}{2\omega_p}\right)} \textrm{.} \label{eq:Z21 weak coupling general result}
\end{equation}
This solution has the required properties of being purely imaginary and reciprocal such that $Z_{12}(\omega)=Z_{21}(\omega)$. This approach can also be applied to find tractable solutions in the case where one of the resonators is mirrored through a line bisecting the MTL section, as well as in the cases where one or both of the resonators are replaced with $\lambda/2$ resonators. Furthermore, under the weak-coupling and consonant-line assumptions, the solution generalizes to resonators that are connected by multiple MTL sections. In this case, the transfer impedance is found by summing the contributions coming from each MTL section treated independently, as depicted in Fig.~\ref{fig:transfer_impedance_multiple_MTLs}. In this way, Eq.~\eqref{eq:Z21 weak coupling general result} can also be used for modeling the response and predicting notch frequencies for resonators that are coupled by multiple MTL sections.
\subsection{Capacitive coupling}
\indent As stated in the main text, lumped-element capacitive coupling represents a special case of the MTL coupling found by taking the limits $c_m \ell_c \to C_J$ and $l_m, \ell_c \to 0$. In this case, the solution for $Z_{21}(\omega)$ given in Eq.~{\eqref{eq:Z21 weak coupling general result}} reduces to
\begin{equation}
    Z_{21}(\omega) = -iZ_0^2 \frac{\sin \! \left( \frac{\omega \ell_r^s}{v}\right) \sin \! \left( \frac{\omega \ell_p^s}{v}\right)}{\cos  \! \left( \frac{\pi\omega}{2\omega_r} \right) \cos \! \left(\frac{\pi\omega}{2\omega_p}\right)} \omega C_J \textrm{.}
    \label{eq:Z21 for direct capacitive coupling}
\end{equation}
This expression has its first zeros at frequencies $\pi v/\ell_{r}^s$ and $\pi v/\ell_{p}^s$. These terms satisfy $\pi v/\ell_{r}^s \geq 2\omega_{r}$ and $\pi v/\ell_{p}^s \geq 2\omega_{p}$, leading to the bound on the notch frequency $\omega_n \geq \textrm{min}(2\omega_{r},2\omega_{p})$ that is stated in the main text.
\subsection{MTL coupling}
\indent
For a two-line, cyclic-symmetric MTL in a homogeneous medium with relative permittivity $\epsilon_r$, the per-length capacitances and inductances are constrained to obey to relations~\cite{paul2007analysis}
\begin{align}
& l_c(c_c+c_m)-l_mc_m=\mu_0\epsilon_0\epsilon_r \textrm{,} \\
& Z_m=Z_c \textrm{.}
\label{eq:homogeneous identity two}
\end{align}
Substituting Eq.~\eqref{eq:homogeneous identity two} into Eq.~\eqref{eq:Z21 weak coupling general result} and using the consonant-line assumption to set $Z_c=Z_0$, the solution for $Z_{21}(\omega)$ reduces to 
\begin{figure}
\includegraphics[width=0.475\textwidth,height=0.4\textheight,keepaspectratio]{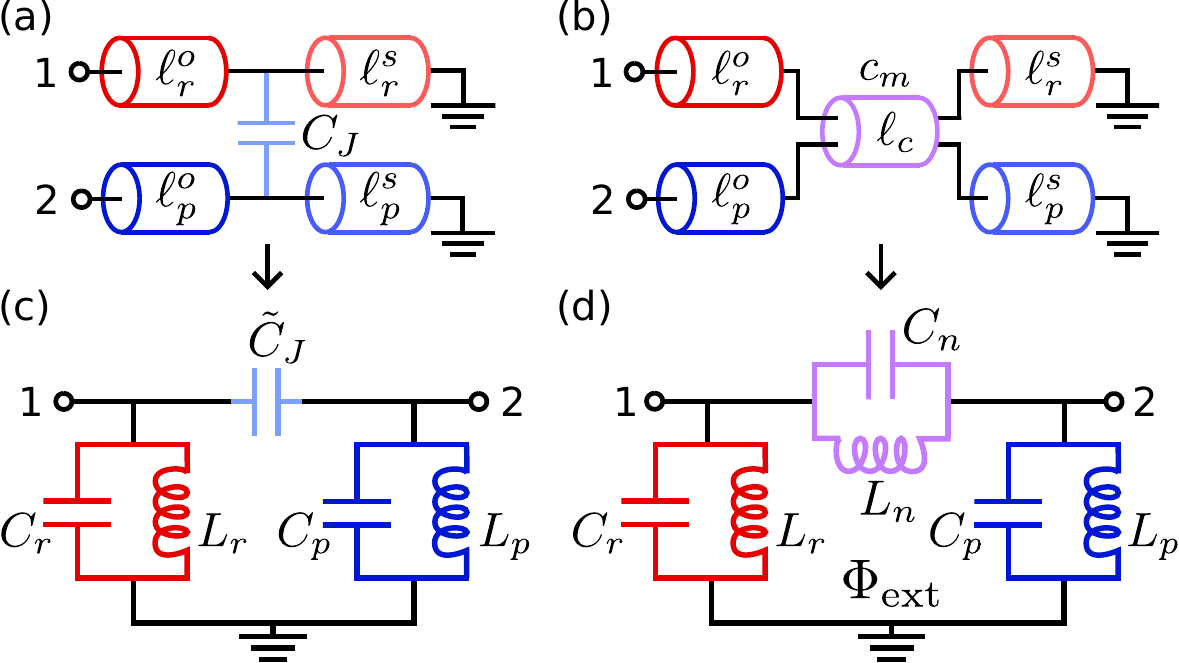}
\caption{\label{fig:distributed and equivalent circuits - appendix} Distributed circuits and corresponding equivalent lumped-element circuits for the coupled readout and filter resonators. The equivalent-circuit models are used to determine the coupling strength between the $\lambda/4$ modes of the readout resonator and filter resonator, for the cases of (a)[(c)]~lumped-element capacitive coupling, and (b)[(d)]~MTL coupling.}
\end{figure}
\begin{equation}
    Z_{21}(\omega) = iZ_0 \frac{\sin \! \left( \frac{\omega \ell_c}{v} \right) \cos \! \left( \frac{\pi \omega}{2\omega_n}\right)} {\cos \! \left( \frac{\pi\omega}{2\omega_r} \right) \cos \! \left(\frac{\pi\omega}{2\omega_p} \right)} \left(\frac{c_m}{c} \right) \textrm{.}
\label{eq:Z21 for homogeneous coupling - appendix}
\end{equation}
Here, $\omega_n$ is the notch frequency where $Z_{21}(\omega)=0$, which we restate takes the form
\begin{equation}
\omega_n= \frac{\pi v}{ 2\left(\ell_r^s+\ell_c+\ell_p^s\right)} \mathrm{.}
\label{eq:notch frequency - appendix}
\end{equation}

%% file: appendix_body/Supplementary_C.tex
\section{Equivalent lumped-element circuits and interaction Hamiltonians}
\label{appendix:Deriving equivalent circuits and Interaction Hamiltonian}
In this section, we provide a mapping between the distributed and equivalent lumped-element circuits shown in Fig.~\ref{fig:distributed and equivalent circuits - appendix}. The mapping is chosen such that the equivalent circuits closely match the electrical response of the distributed circuits around the readout- and filter-resonator frequencies. The calculated transfer impedances for the distributed and equivalent circuits are shown in Fig.~\ref{fig:equivalent circuits}(c) of the main text for the particular set of circuit parameters given in Table~\ref{tab:resonator dimensions - appendix}, indicating good agreement. Given the electrical responses closely match, we infer that the interaction Hamiltonians for both the distributed and equivalent circuits are also closely matched.
\\ 
\indent We first review that a $\lambda/4$ transmission-line resonator with length $\ell$ has its fundamental resonance at $\omega_{\lambda/4} = \pi v/(2\ell)$ and that at frequencies near this resonance the electrical response is well approximated by a parallel LC resonator with lumped-element values given by~\cite{Pozar2011}
\begin{equation}
C_{\lambda/4} = \frac{c \ell}{2} \label{eq:C and L equation for lambda/4 resonator - in terms of c}  \textrm{,} \quad \quad 
L_{\lambda/4} = \frac{8l \ell}{\pi^2 }  \textrm{,} 
\end{equation}
where $c$ and $l$ are the capacitance and inductance per length of the line, respectively, satisfying $c=1/(Z_0 v)$ and $l=Z_0/v$. Using this result, the lumped-element capacitance and inductance 
of the readout resonator and filter resonator are mapped from the distributed circuit by the relations,
\begin{equation}
C_{r(p)} = \frac{\ell_{r(p)}}{2Z_0v}   \textrm{,}  \quad
L_{r(p)} = \frac{8Z_0 \ell_{r(p)}}{\pi^2v}   \textrm{.} \quad
\label{eq:C and L for lambda/4 resonator - appendix}
\end{equation}
The characteristic impedances of the lumped-element resonators, defined $Z_{r(p)}\equiv \sqrt{L_{r(p)}/C_{r(p)}}$, are related to the transmission-line characteristic impedance by the formula
\begin{equation}
Z_{r(p)}  = \frac{4Z_0}{\pi}  \textrm{.}
\label{eq:Characteristic impedance equation for lambda/4 resonator - appendix}
\end{equation}
\begin{table}
\centering
\caption{\label{tab:resonator dimensions - appendix} Readout- and filter-resonator dimensions used to calculate the transfer impedances plotted in Fig.~\ref{fig:equivalent circuits}(c) of the main text. The Cap(MTL) parameters were used to calculate the transfer impedance for the distributed circuit with the lumped-element coupling capacitor(MTL coupler). The value of $C_J$ was set to \SI{1.4}{\femto\farad}, and the value of $c_m$ was set to $\SI{8.5}{\femto\farad / \milli\meter}$ in order to match the coupling strengths. Values of $v=1.19\times10^8\si{\meter/\second}$ and $Z_0=\SI{66}{\ohm}$ were used for the phase velocity and characteristic impedance of the lines. The MTL parameters match the designed parameters for the readout and filter resonators of Q$_1$.}
\begin{ruledtabular}
\begin{tabular}{ccccccccc}
& $\ell_r^o$ & $\ell_r^s$ &  $\ell_p^o$ & $\ell_p^s$ &  $\ell_c$ & $d$ \\
 & (\si{\micro\meter}) & (\si{\micro\meter}) & (\si{\micro\meter}) & (\si{\micro\meter}) & (\si{\micro\meter}) & (\si{\micro\meter}) \\
\hline
Cap & $1133$ & $1776$ & $918$ & $1818$ & $\textrm{--}$ & $\textrm{--}$ \\
MTL & $974$ & $1617$ & $759$ & $1659$ & $318$ & $5.5$\\
\end{tabular}
\end{ruledtabular}
\end{table}
\subsection{Capacitive coupling}
Figure~\ref{fig:distributed and equivalent circuits - appendix}(c) depicts the equivalent circuit for the capacitively coupled $\lambda/4$ resonators. Equating the transfer impedance $Z_{21}^\mathrm{ec}(\omega)$ for the equivalent circuit with the transfer impedance for the distributed circuit in Fig.~\ref{fig:distributed and equivalent circuits - appendix}(a) [given by Eq.~\eqref{eq:Z21 for direct capacitive coupling}] yields the following expression for the ratio between the equivalent-circuit capacitance $\tilde{C}_J$ and the distributed-circuit capacitance $C_J$,
\begin{equation}
\frac{\tilde{C}_J}{C_J} = \frac{\pi^2}{16} \frac{\left(\frac{\omega_r}{\omega}-\frac{\omega}{\omega_r}\right)\left(\frac{\omega_p}{\omega}-\frac{\omega}{\omega_p}\right)}{\cos \!\left(\frac{\omega \pi}{2\omega_r}\right)\cos \!\left(\frac{\omega \pi}{2\omega_p}\right)} \sin\! \left(\frac{\ell_r^s\omega}{v}\right)\sin \!\left(\frac{\ell_p^s\omega}{v}\right)  \textrm{,} 
\end{equation}
where the weak-coupling assumption $\tilde{C}_J \ll C_r, C_p$ has been used. This ratio is frequency-dependent. However the frequency dependence is weak around the the $\lambda/4$-mode frequencies $\omega_r$ and $\omega_p$. In order to determine a frequency-independent value for $\tilde{C}_J$ that is accurate around these modes, the expression is separated into readout-resonator- and filter-resonator-dependent factors, and these are then evaluated with $\omega$ set to the corresponding resonance frequencies 
\begin{align}
& \frac{\tilde{C}_J}{C_J}  \approx \left(\frac{\pi}{4}\right)^{\! 2} A_r A_p  \textrm{,} \\
& A_r = \left[\frac{\left(\frac{\omega_r}{\omega}-\frac{\omega}{\omega_r}\right)}{\cos{\left(\frac{\omega \pi}{2\omega_r}\right)}}\sin{\left(\frac{\ell_r^s\omega}{v}\right)}\right]\Bigg\rvert_{\omega \to \omega_r}   \textrm{,} \\
& A_p = \left[\frac{\left(\frac{\omega_p}{\omega}-\frac{\omega}{\omega_p}\right)}{\cos{\left(\frac{\omega \pi}{2\omega_p}\right)}} \sin{\left(\frac{\ell_p^s\omega}{v}\right)}\right]\Bigg\rvert_{\omega \to \omega_p}  \textrm{.} 
\end{align}
Evaluating the factors yields
\begin{equation}
\tilde{C}_J \approx  C_J \sin\! \left(\frac{\ell_r^s\omega_r}{v}\right)\sin \! \left(\frac{\ell_p^s\omega_p}{v}\right) \textrm{.} 
\label{eq:equivalent circuit direct capacitance value}
\end{equation}
According to this approximation, the equivalent-circuit capacitance vanishes as the distributed-circuit capacitor approaches the shorted end of either $\lambda/4$ resonator and takes its maximum value when the distributed-circuit capacitor is connected to the open end of both resonators, in which case $\tilde{C}_J = C_J$. \\
\indent The equivalent lumped-element circuit in Fig.~\ref{fig:distributed and equivalent circuits - appendix}(c) is solved by the branch-flux method~\cite{devoret1995quantum, vool2017introduction} to yield the following interaction Hamiltonian between the readout-resonator and dedicated-filter modes under the rotating-wave approximation,
\begin{align}
     & \hatH_{\mathrm{int}} = \hbar J^{\mathrm{cap}} \left(\hat{a}_r^\dag \hat{a}_p + \hat{a}_r \hat{a}_p^\dag \right)  \textrm{,}  \\
    &     J^{\mathrm{cap}} = \frac{1}{2}\sqrt{Z_r Z_p}\, \omega_r \omega_p\tilde{C}_J  \textrm{.} \label{eq:J_coupling direct capacitive - appendix}
\end{align}
The solution is valid assuming weak coupling between the modes such that $\tilde{C}_J \ll C_r, C_p$. Substituting in Eq.~\eqref{eq:Characteristic impedance equation for lambda/4 resonator - appendix} for the characteristic impedances $Z_{r(p)}$ and Eq.~\eqref{eq:equivalent circuit direct capacitance value} for the capacitance $\tilde{C}_J$ yields
\begin{equation}
J^{\mathrm{cap}} = \frac{2}{\pi} Z_0 \omega_r \omega_p C_J \sin \! \left( \frac{\omega_r \ell_r^s}{v}\right) \sin \! \left( \frac{\omega_p \ell_p^s}{v}\right) \textrm{.}
\label{eq2:J_coupling direct capacitive}
\end{equation}
This expression gives the coupling strength $J^{\mathrm{cap}}$ purely in terms of the distributed-circuit parameters.
\subsection{MTL coupling}
Figure~\ref{fig:distributed and equivalent circuits - appendix}(d) depicts the equivalent circuit for the MTL-coupled $\lambda/4$ resonators. The form of the circuit is motivated as the coupling capacitance $C_n$ and coupling inductance $L_n$ destructively interfere, resulting in a notch frequency where transmission through the coupled resonators is suppressed. 
To determine a mapping for the coupling elements $C_n$ and $L_n$, we match the transfer impedance $Z_{21}^{\mathrm{ec}}$ for the equivalent circuit with the transfer impedance $Z_{21}$ for the distributed circuit in Fig.~\ref{fig:distributed and equivalent circuits - appendix}(b) [given by Eq.~\eqref{eq:Z21 for homogeneous coupling - appendix}] around the notch frequency $\omega_n$. We equate both the zeroth- and first-order derivatives,
\begin{align}
& Z_{21}^{\mathrm{ec}}(\omega_n) = Z_{21}(\omega_n) = 0 \textrm{,}\\
& \frac{d Z_{21}^{\mathrm{ec}}(\omega)}{d \omega}\bigg\rvert_{\omega = \omega_n} = \frac{d Z_{21}(\omega)}{d \omega}\bigg\rvert_{\omega = \omega_n} \textrm{.}
\label{eq:blablabla}
\end{align}
The first equation gives the condition $\sqrt{L_nC_n}=1/\omega_n$, and the second equation gives the condition
\begin{equation}
Z_n \equiv \sqrt{\frac{L_n}{C_n}} = Z_0 \frac{64}{\pi^3} \frac{\cos \! \left(\frac{\pi \omega_n}{2 \omega_r} \right)\cos \! \left(\frac{\pi \omega_n}{2 \omega_p}\right)}{\left( \frac{\omega_r}{\omega_n} - \frac{\omega_n}{\omega_r} \right) \left( \frac{\omega_p}{\omega_n} - \frac{\omega_n}{\omega_p} \right)}  \frac{\left(c/c_m\right)}{\sin \! \left( \frac{\omega_n \ell_c}{v} \right)}  \textrm{.}  
\label{eq:notch characteristic impedance symbolic}
\end{equation}
\indent The effective Hamiltonian of the MTL-coupled resonators can now be found because the Hamiltonian of the equivalent lumped-element circuit can be constructed following a standard procedure~\cite{devoret1995quantum, vool2017introduction}. 
The external flux $\Phi_{\textrm{ext}}$ through the inductive loop is set to zero, as the distributed circuit from which the equivalent circuit is constructed has no such inductive loop. Focusing on the interaction part of the Hamiltonian and making the RWA yields 
\begin{align}
     & \hatH_{\mathrm{int}} =  \hbar J \left(\hat{a}_r^\dag \hat{a}_p + \hat{a}_r \hat{a}_p^\dag \right) \textrm{,} \\
     & J = \frac{\sqrt{Z_r Z_p}}{2 Z_{n}}\sqrt{\omega_r\omega_p}\left(\frac{\sqrt{\omega_r\omega_p}}{\omega_n} - \frac{\omega_n}{\sqrt{\omega_r\omega_p}}\right)  \textrm{.} 
     \label{eq:MTL coupling J - appendix}
\end{align}
The solution is valid assuming weak coupling between the modes such that $ C_n \ll C_r, C_p$ and $L_n \gg L_{r}, L_p$. The solutions to $Z_{r(p)}$ and $Z_{n}$ given in Eqs.~\eqref{eq:Characteristic impedance equation for lambda/4 resonator - appendix} and \eqref{eq:notch characteristic impedance symbolic} respectively are substituted into this expression. In addition, we make the change of variables $\omega_r = \overline{\omega}_{rp} + \Delta_{rp}/2$ and $\omega_p = \overline{\omega}_{rp} - \Delta_{rp}/2$, where $\overline{\omega}_{rp}$ is the average of the resonator and filter frequencies and $\Delta_{rp}$ is the detuning. Expanding in orders of the term $\Delta_{rp} / \omega_n$ yields
\begin{multline}
J = \overline{\omega}_{rp} \frac{\pi^2}{32} \frac{\left(\frac{\overline{\omega}_{rp}}{\omega_n} - \frac{\omega_n}{\overline{\omega}_{rp}}\right)^{\! 3}}{\cos^2 \! \left(\frac{\pi\omega_n}{2 \overline{\omega}_{rp}}\right)} \left(\frac{c_m}{c} \right) 
\sin \! \left(\frac{\omega_n\ell_c}{v}\right) \\ + \mathcal{O} \left[ \left( \frac{\Delta_{rp}}{\omega_n} \right)^{\!\! 2} \right]  \textrm{.} 
\label{eq:J_coupling - appendix}
\end{multline}
For readout applications it is typically desirable to make the detuning $\Delta_{rp}$ as small as possible and so the higher order terms can be neglected. This results in Eq.~\eqref{eq2:J_coupling} in the main text, which gives the exchange coupling strength between two MTL-coupled $\lambda/4$ resonators in terms of the resonator frequencies, the notch frequency, and the length and coupling capacitance of the MTL-coupler. 

%% file: appendix_body/Supplementary_D.tex
\section{Circuit models for the Purcell-limited relaxation time}
\label{appendix:Purcell filtering enhancement expressions}
\begin{figure}
\includegraphics[width=0.325\textwidth,height=0.4\textheight,keepaspectratio]{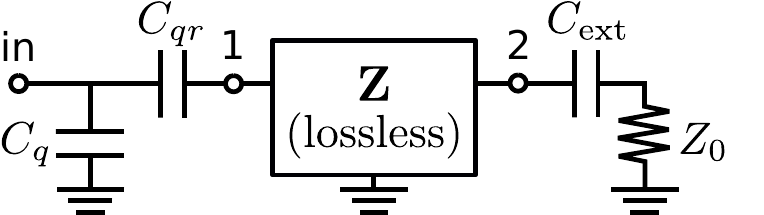}
\caption{\label{fig:T1 limit solving circuit} Circuit model for determining the Purcell-limited relaxation time of a qubit. A qubit with shunt capacitance $C_q$ is capacitively coupled to port 1 of a lossless 2-port network with impedance matrix $\boldsymbol{\mathrm{Z}}$. Port 2 of the network is capacitively coupled to a load with resistance $Z_{0}$.}
\end{figure}

In this section, the circuit models used to predict the Purcell-limited qubit relaxation time $T_1^{\mathrm{pl}}$ are described. The models are then used to derive an expression for the Purcell-filtering enhancement provided by the intrinsic notch filter.
\\ 
\indent Figure~\ref{fig:T1 limit solving circuit} shows the general circuit used to model the energy relaxation of the qubit due to its coupling to the readout line. The qubit, with shunt capacitance $C_q$, is coupled via a capacitor with capacitance $C_{qr}$ to a lossless 2-port network. The network is coupled at port 2 to a readout line with characteristic impedance $Z_{0}$ via a capacitor with capacitance $C_{\textrm{ext}}$.
The characteristic relaxation time of the qubit imposed by its coupling to the readout line can be modeled by the following equation~\cite{houck2008controlling},
\begin{equation}
    T_1^\mathrm{pl}(\omega_q) = \frac{C_q}{\text{Re}[Y_{\textrm{in}}(\omega_q)]}  \textrm{.} 
    \label{eq:Houck_T1}
\end{equation}
Defining $Z_{11}$, $Z_{22}$, and $Z_{21}$ as the impedance matrix elements for the lossless 2-port network, the real part of the input admittance $Y_{\mathrm{in}}$ for the circuit in Fig.~\ref{fig:T1 limit solving circuit} is given by
\begin{align}
    & \text{Re}[Y_{\textrm{in}}(\omega)] = \frac{Z_0 \left|Z_{21}\right|^2}{\left|\left(Z_\textrm{11} + Z_{qr}\right)\left(Z_{22}+Z_\textrm{ext} + Z_0 \right)\right|^2} \label{eq:real admittance expression}  \textrm{,}  \\
    & Z_{qr} = \frac{-i}{\omega C_{qr}}  \textrm{,} \quad Z_{\textrm{ext}} = \frac{-i}{\omega C_{\textrm{ext}}}  \textrm{.} 
\end{align}
This expression is valid when ${Z_{21} \ll Z_{11}, Z_{22}}$. The transfer impedances $Z_{21}$ for the 2-port network given the specific circuits shown in Figs.~\ref{fig:distributed and equivalent circuits - appendix}(c) and (d) are defined as $Z_{21}^{\textrm{cap}}$ and $Z_{21}^{\textrm{MTL}}$, respectively. Likewise, the Purcell-limited relaxation times of the qubit given these specific circuits are defined as $T_1^{\textrm{cap}}$ and $T_1^{\textrm{MTL}}$. 
\\
\indent The predicted Purcell-limited relaxation times plotted in Fig.~\ref{fig:fitted_spectroscopy_experiments}(b) of the main text were generated using $T_1^{\textrm{cap}}$ and $T_1^{\textrm{MTL}}$. The circuit parameters were constrained to match the measured frequencies, coupling strengths and external linewidths for the qubit and the readout and filter resonators, in a similar manner to Ref.~\citenum{yen2024interferometric}. For completeness, the shunting impedance to ground given by Eq.~\eqref{eq:shunt impedance - appendix} in Appendix E was included in the model. The effect of the shunt impedance is the following adjustment to the values of $Z_\mathrm{ext}$ and $Z_0$,
\begin{align}
    & Z_\mathrm{ext} \to Z_\mathrm{ext} +\frac{Z_\mathrm{shunt}}{1 + \left|Z_\mathrm{shunt}/Z_0\right|^2} \mathrm{,} \\ 
    & Z_0 \to \frac{Z_0 }{1 + \left|Z_0/Z_\mathrm{shunt}\right|^2} \mathrm{.}
\end{align}
Including the shunt impedance in the model results in a modest increase in the Purcell-limited relaxation times at the qubit frequencies of around $5$ to $\SI{10}{\percent}$.
\\
\indent For weak coupling between the readout resonator and dedicator filter and at frequencies away from their resonances, $Z_{11}$ and $Z_{22}$ are to a good approximation independent of the coupling-circuit implementation such that $Z_{11}^{\textrm{cap}}=Z_{11}^{\textrm{MTL}}$ and $Z_{22}^{\textrm{cap}}=Z_{22}^{\textrm{MTL}}$. As a result, the ratio of the relaxation times takes the form
\begin{equation}
    \frac{T_1^{\textrm{MTL}}\left(\omega_q\right)}{T_1^{\textrm{cap}}\left(\omega_q\right)} = \left|\frac{Z_{21}^{\textrm{cap}}\left(\omega_q\right)}{Z_{21}^{\textrm{MTL}}\left(\omega_q\right)}\right|^{ 2}  \textrm{.} 
\end{equation}
Given that the readout resonator and dedicated filter are weakly coupled, this reduces to
\begin{equation}
    \frac{T_1^{\textrm{MTL}}\left(\omega_q\right)}{T_1^{\textrm{cap}}\left(\omega_q\right)}  = \left[\frac{\omega_q Z_{n} \tilde{C}_J}{\left(\frac{\omega_n}{\omega_q} - \frac{\omega_q}{\omega_n}\right)}\right]^{\!2}  \textrm{.} \label{eq:T1 ratios}
\end{equation}
Using Eqs.~(\ref{eq:J_coupling direct capacitive - appendix}) and~\eqref{eq:MTL coupling J - appendix} from Appendix C, the parameters $\tilde{C}_J$ and $Z_{n}$ can be expressed in terms of the coupling strengths $J^{\mathrm{cap}}$ and $J^{\mathrm{MTL}}$, respectively, yielding
\begin{align}
    & \tilde{C}_J  = \frac{\pi J^{\textrm{cap}}}{2Z_0\overline{\omega}_{rp}^2}   + \mathcal{O}\left[\left(\frac{\Delta_{rp}}{\overline{\omega}_{rp}}\right)^{\!\!2}\right] \textrm{,} \\
    & Z_{n} = \frac{2}{\pi} Z_0\left(\frac{\overline{\omega}_{rp}}{\omega_n} - \frac{\omega_n}{\overline{\omega}_{rp}}\right)\frac{\overline{\omega}_{rp}}{J^{\textrm{MTL}}}  + \mathcal{O}\left [ \left(\frac{\Delta_{rp}}{\omega_n}\right)^{\!\!2} \right] \textrm{.} 
\end{align}
Dropping the higher-order terms and substituting these expressions into Eq.~\eqref{eq:T1 ratios} expresses the ratio of the relaxation times in terms of frequencies and coupling strengths. The coupling strengths $J^{\textrm{cap}}$ and $J^{\textrm{MTL}}$ are then equated so as to ensure a fair comparison between the $T_1$ limits, yielding
\begin{equation}
    \frac{T_1^{\textrm{MTL}}\left(\omega_q\right)}{T_1^{\textrm{cap}}\left(\omega_q\right)} = \frac{1}{4} \frac{\omega_q^2}{\Delta_{qn}^2}\left(1-\frac{\omega_n^2}{\overline{\omega}_{rp}^2} \right) ^{\!2} \left(1 + \mathcal{O}\left [\left(\frac{\Delta_{qn}}{\omega_n}\right)^{\!\!2}\right]\right)  \textrm{,}
\end{equation}
where $\Delta_{qn} = \omega_q - \omega_n$ is the detuning of the qubit from the notch. Dropping the higher-order terms and defining ${\xi \equiv T_1^{\textrm{MTL}}\left(\omega_q\right)/T_1^{\textrm{cap}}\left(\omega_q\right)}$ results in Eq.~\eqref{eq:Notch filter enhancement factor} in the main text.

%% file: appendix_body/Supplementary_E.tex
\section{Semi-classical model for the readout system dynamics}
\label{appendix:Semi-classical model for the readout system dynamics}
\begin{figure}
\includegraphics[width=0.435\textwidth,height=0.475\textheight,keepaspectratio]{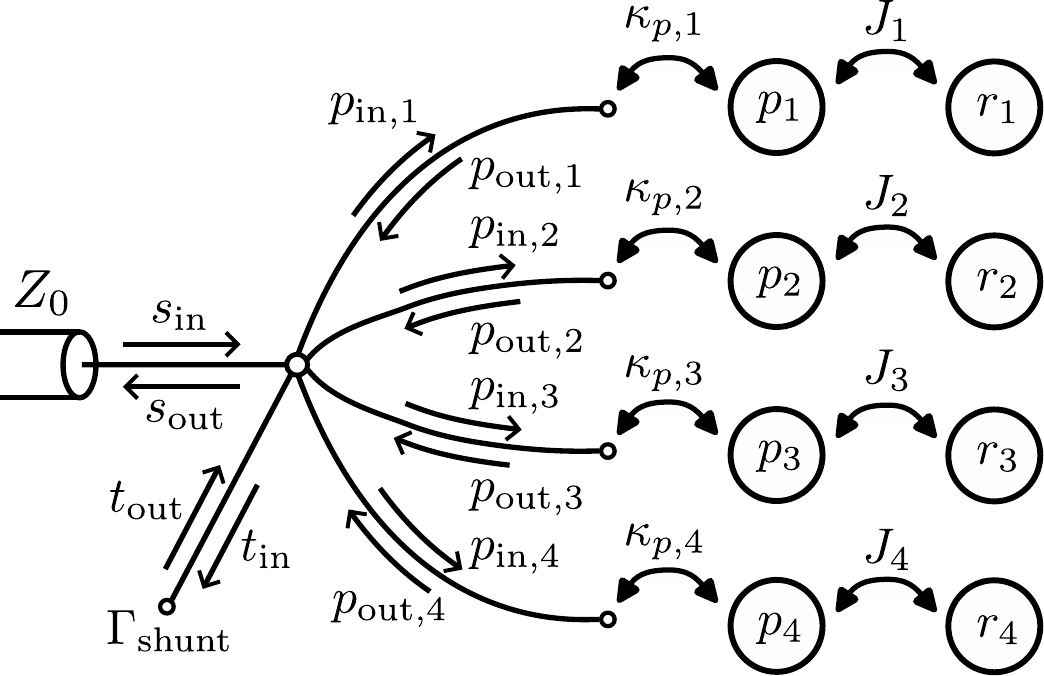}
\caption{\label{fig:input output multiplexed reflection model} Input--output network for the multiplexed system of four readout resonators of amplitudes $r_j$ and dedicated filter resonators of amplitudes $p_j$~($ j=1,\ldots,4$). The multiplexed resonators couple to the end of a transmission line with characteristic impedance $Z_0$. The model is generalized to include a shunt impedance to ground.}
\end{figure}

This section describes the semi-classical model for the multiplexed readout and filter resonators. The model is utilized to predict both the steady-state and time-dependent responses of the readout system to an external drive, and to determine the system's normal modes.
\\
\indent Figure~\ref{fig:input output multiplexed reflection model} shows the input--output network used to model the multiplexed readout system. A traveling coherent field $s_{\textrm{in}}$ propagating along a transmission line with characteristic impedance $Z_0$ is incident on the system of coupled readout and filter resonators. The traveling fields are treated classically when considering their behavior at the shunt and at the node where the fields scatter. They are treated using input--output theory~\cite{gardiner1985input} when considering their interaction with the resonators. We denote the coherent field amplitudes of filter resonator~$j$ and readout resonator $j$ as $p_j$ and $r_j$, respectively. The term $\kappa_{p,j}$ is the external linewidth of filter mode~$j$, and the term $J_j$ is the coupling strength between filter resonator~$j$ and readout resonator~$j$. The model is linear, and qubits are treated as only inducing a state-dependent shift to the readout-resonator frequencies.
\\ 
\indent The reflection coefficient of the incident field is defined by 
\begin{equation}
\Gamma_{\textrm{incident}} \equiv \frac{s_\textrm{out}}{s_\textrm{in}} \textrm{.} \label{eq:incident reflection coeff - appendix}
\end{equation}
The incident field is treated as separating into distinct fields that each interact with a single filter mode or with the shunt. The incoming and outgoing fields at the node where the separation occurs obey the relations
\begin{align}
& s_\textrm{out} - s_{\textrm{in}} = t_\textrm{out} - t_\textrm{in} + \sum_{j=1}^{4} \left(p_{\textrm{out},j} - p_{\textrm{in},j}\right) \textrm{,} \label{eq:energy conservation - appendix} \\
& s_{\textrm{out}} + s_{\textrm{in}} = t_\textrm{out} + t_\textrm{in} = p_{\textrm{out},j} + p_{\textrm{in},j} \textrm{.} \label{eq:field continuity 1 - appendix} 
\end{align}
These conditions result from Kirchhoff's laws. The fields incident at the shunt and the filter resonators are reflected with reflection coefficients defined by
\begin{equation}
    \Gamma_{\textrm{shunt}} \equiv \frac{t_\textrm{out}}{t_\textrm{in}}  \textrm{,} \quad
    \Gamma_{p,j} \equiv \frac{p_{\textrm{out},j}}{p_{\textrm{in},j}} \textrm{.} \label{eq:shunt and dedicated filter mode reflection coefficients}
\end{equation}
Combining Eqs.~\eqref{eq:incident reflection coeff - appendix}--\eqref{eq:shunt and dedicated filter mode reflection coefficients} yields the following solution for the reflection coefficient of an incident readout signal,
\begin{equation}
\frac{1-\Gamma_{\textrm{incident}}}{1+\Gamma_{\textrm{incident}}} = \frac{1-\Gamma_{\textrm{shunt}}}{1+\Gamma_{\textrm{shunt}}} + \sum_{j=1}^4\left(\frac{1-\Gamma_{p,j} }{1+\Gamma_{p,j}}\right) \textrm{.}
\label{eq:incident reflection - appendix}
\end{equation}
We note that this expression is equivalent to stating that the total load impedance terminating the transmission line is the parallel sum of the shunt impedance and the readout-circuit impedance. \\
\indent The reflection coefficient at the shunt is given by 
\begin{equation}
    \Gamma_{\textrm{shunt}} = \frac{Z_{\textrm{shunt}} - Z_0}{Z_{\textrm{shunt}} + Z_0} \textrm{.}
    \label{eq:shunt reflection}
\end{equation}
Here, the shunt impedance $Z_{\textrm{shunt}}$ takes the form of a parallel LC resonator
\begin{equation}
\frac{1}{Z_{\textrm{shunt}}} = -\frac{i}{\omega L_{\textrm{shunt}}} + i \omega C_{\textrm{shunt}} \textrm{.} 
\label{eq:shunt impedance - appendix}
\end{equation}
The term $C_{\textrm{shunt}}$ is the parasitic shunt capacitance of the TSV structure that routes signals from the pogo pin on the circuit backside to the filter resonators, and $L_{\textrm{shunt}}$ is due to the backside spiral inductance that connects the TSV structure to the ground plane~[Figs.~4(e) and~A1(c)]. Shunt values of $C_{\textrm{shunt}}=\SI{230}{\femto\farad}$ and $L_{\textrm{shunt}}=\SI{1.01}{\nano\henry}$ were determined by COMSOL simulations and an analytic formula for the inductance of a spiral inductor~\cite{mohan1999simple}. At the frequency $1/\sqrt{L_\textrm{shunt}C_\textrm{shunt}}\,=\SI{10.44}{\giga\hertz}$, the shunt impedance goes to infinity and thus the reflection coefficient $\Gamma_{\mathrm{shunt}}$ approaches $1$. As a result, around the readout frequencies, the effect of the large shunt capacitance is screened out by the shunt inductance.
\indent Using input--output theory~\cite{gardiner1985input}, the incident and reflected fields at the filter resonators are related to the coherent field amplitudes of the filter resonators by the expression
\begin{equation}
p_{\textrm{out},j} = p_{\textrm{in},j} - \sqrt{\kappa_{p,j}}\, p_j \textrm{.} \label{eq:incoming outgoing field relation - appendix} 
\end{equation}
The equations of motion for the coupled filter-mode and readout-resonator-mode system are given by
\begin{align}
& \frac{d p_{j}}{dt} = -i \Delta_{pd,j\,}p_j - \frac{\kappa_{p,j} + \gamma_{p,j}}{2}p_j - iJ_j r_j + \sqrt{\kappa_{p,j}}\, p_{\textrm{in},j}\textrm{,} \label{eq:filter mode dynamics} \\
& \frac{d r_{j}}{dt} = -i \Delta_{rd,j\,}^{g(e)}r_j - \frac{\gamma_{r,j}}{2}r_j - iJ_j p_j \label{eq:readout mode dynamics} \textrm{.}
\end{align}
Here, $\Delta_{pd,j}=\omega_{p,j} - \omega_d$ is the detuning of filter mode~$j$ from the drive, $\Delta_{rd,j}^{g(e)}=\omega_{r,j}^{g(e)} - \omega_d$ is the qubit-state-dependent detuning of readout mode $j$ from the drive, and $\omega_{r,j}^{e} = \omega_{r,j}^{g} + 2\chi_j$ is the readout-mode frequency given the qubit is in the $\ket{e}$ state. The terms $\gamma_{r,j}$ and $\gamma_{p,j}$ are the internal linewidths of readout resonator $j$ and filter resonator $j$, respectively.
\subsection{Steady-state results}
\indent Solving the equations of motion in the steady state and combining with Eq.~\eqref{eq:incoming outgoing field relation - appendix} yields the following solution for the qubit-state-dependent reflection coefficients at the dedicated filters 
\begin{equation}
\Gamma_{p,j} ^{g(e)} = 1 - \frac{4i\kappa_{p,j}\left(\Delta_{rd,j}^{g(e)} - i\gamma_{r,j}/2\right)}{\left(2i \Delta_{pd,j} + \kappa_{p,j} + \gamma_{p,j}\right)\left(2i \Delta_{rd,j}^{g(e)} + \gamma_{r,j}\right) + 4J_j^2} \textrm{.} \label{eq: Gamma_j factor result}
\end{equation}
By combining this result with Eq.~\eqref{eq:shunt reflection} for the reflection coefficient at the shunt, the steady-state reflection at the device given by Eq.~\eqref{eq:incident reflection - appendix} is fully determined from the readout and shunt parameters. An important detail is
that input–output theory uses the clockwise sign convention for phasor rotation whereas Eq.~\eqref{eq:shunt impedance - appendix} for the shunt impedance uses the counterclockwise convention from electrical engineering. When fitting to the measured data we choose the electrical engineering convention and so all values of $i$ in Eq.~\eqref{eq: Gamma_j factor result} are replaced with $-i$. The measured reflection response $\Gamma_{\textrm{meas}}$ was fit to the model function
\begin{equation}
\textrm{arg}\left(\Gamma_{\textrm{meas}} \right) = \textrm{arg}\left( \Gamma_{\textrm{incident}} \right) + \theta_0 - \omega \tau \textrm{.} 
\label{eq:reflected phase model}
\end{equation}
Here, $\theta_0$ is a constant phase offset, and $\tau$ accounts for the electrical length of the experimental setup. The reflection spectrum $\Gamma_{\textrm{meas}}$ was measured twice over a frequency range that covered all of the readout-resonator and filter-resonator modes; first with all qubits prepared in the $\ket{g}$ state and then with all qubits prepared in the $\ket{e}$ state. The two datasets were then fit simultaneously to extract the parameters $\omega_{r,j}$, $\omega_{p,j}$, $\kappa_{p,j}$, $J_j$ and $\chi_j$ for all the filter resonators and readout resonators. The parameters $\gamma_{p,j}$ and $\gamma_{r,j}$ were set to zero in the model function since the external decay rates of the filter resonators dominated over all the internal decay rates. The bare readout-resonator and filter-resonator parameters extracted from the fit are presented in Table~\ref{tab:Bare readout parameters - appendix}.
\\
\indent The bounds on the noise-photon populations $n_{r,j}^\mathrm{noise}$ of readout-resonators given in the main text were determined from the fitted parameters using the relation~\cite{sunada2024photon}
\begin{equation}
\Gamma_{\phi,j}^{\mathrm{noise}} = \frac{n_{r,j}^\mathrm{noise}}{2} \int_{-\infty}^\infty \left| \Gamma_{\mathrm{incident}}^e(\omega) - \Gamma_{\mathrm{incident}}^g(\omega) \right|^2\frac{d\omega}{2\pi} \mathrm{,}
\end{equation}
where $\Gamma_{\phi,j}^{\mathrm{noise}}$ is the noise-photon-induced dephasing rate in qubit $j$ and $\Gamma_{\mathrm{incident}}^{g(e)}(\omega)$ is the reflection coefficient at the device when qubit $j$ is in the $|g\rangle$ and $|e\rangle$ states, respectively.
\subsection{Time-dependent results}
\indent In order to solve the time dependence of the coherent field amplitudes, it is convenient to use a matrix representation. The driven system of coupled readout and filter resonators can be expressed by the matrix equation
\begin{equation}
\label{eq:coupled_resonators_first_order_ode}
\frac{d}{dt}
\begin{pmatrix}
\mathbf{p} \\
\mathbf{r}
\end{pmatrix}
= 
-i \begin{pmatrix}
\mathbf{P} & \mathbf{J} \\
\mathbf{J} & \mathbf{R}^{g(e)}
\end{pmatrix}
\begin{pmatrix}
\mathbf{p} \\
\mathbf{r}
\end{pmatrix}
+
\begin{pmatrix}
\mathbf{d} \\
\mathbf{0}
\end{pmatrix}
s_\textrm{in}
\textrm{.}
\end{equation}
Here, $\mathbf{P}$ is a $4 \times 4$ matrix for the filter-resonators, $\mathbf{J}$ is a $4 \times 4$ matrix for the couplings between the filter-resonator and readout-resonator modes, and $\mathbf{R}^{g(e)}$ is a qubit-state-dependent $4 \times 4$ matrix for the readout-resonators. The terms $\mathbf{p}$ and $\mathbf{r}$ are the field amplitudes for the filter resonators and readout resonators, respectively given by
\begin{equation}
\label{eq:p_matrix}
\mathbf{p} = \begin{pmatrix}
p_1 \\
p_2 \\
p_3 \\
p_4 
\end{pmatrix}
\textrm{,} \quad
\mathbf{r} = \begin{pmatrix}
r_1 \\
r_2 \\
r_3 \\
r_4 
\end{pmatrix}
\textrm{,}
\end{equation}
and $\mathbf{d}$ gives the coupling coefficients of the driving field $s_\mathrm{in}$ to the filter resonators,
\begin{equation}
\label{eq:d_vector}
\mathbf{d} = \left(\frac{1+\Gamma_{\textrm{shunt}}}{2}\right) \begin{pmatrix}
\sqrt{\kappa_{p,1}} \\
 \sqrt{\kappa_{p,2}} \\
\sqrt{\kappa_{p,3}} \\
\sqrt{\kappa_{p,4}} 
\end{pmatrix}
\textrm{.}
\end{equation}
The matrix $\mathbf{P}$ has elements given by
\begin{equation}
\label{eq:P_matrix}
\mathbf{P}_{ij} = \left(\Delta_{pd,i} -i\frac{\gamma_{p,i}}{2}\right)\delta_{ij} - i \frac{\sqrt{\kappa_{p,i\,}\kappa_{p,j}}}{4}\left(1 + \Gamma_{\textrm{shunt}}\right) \textrm{,}
\end{equation}
where $\delta_{ij}$ is the Kronecker delta function. The second term introduces off-diagonal elements to the matrix $\mathbf{P}$, which represent coupling between the different filter modes. This is due to the fact that the filter resonators all couple to the readout line via a shared node. The matrix $\mathbf{J}$ is diagonal with elements given by
\begin{equation}
\label{eq:J_matrix}
\mathbf{J}_{ij} = J_i\delta_{ij}\textrm{.}
\end{equation}
This matrix couples filter-resonator amplitude $p_j$ to readout-resonator amplitude $r_j$ with coupling strength $J_j$. Finally, the matrix $\mathbf{R}^{g(e)}$ is diagonal with elements given by
\begin{equation}
\label{eq:R_matrix}
\mathbf{R}_{ij}^{g(e)} = \left(\Delta_{rd,i}^{g(e)} - i\frac{\gamma_{r,i}}{2}\right)\delta_{ij} \textrm{.}
\end{equation}
This matrix contains all of the qubit-state dependence and has 16 different permutations depending on the four-qubit state. The system of coupled first-order differential equations in Eq.~\eqref{eq:coupled_resonators_first_order_ode} can be solved numerically to find the time evolution of the coherent field amplitudes of the readout and filter resonators under a drive. The time-dependent output field $s_\mathrm{out}(t)$ into the transmission line is then determined from the filter-resonator amplitudes through Eqs.~\eqref{eq:energy conservation - appendix},~\eqref{eq:field continuity 1 - appendix} and~\eqref{eq:incoming outgoing field relation - appendix}, yielding
\begin{equation}
    s_\mathrm{out}(t) = \Gamma_\mathrm{shunt}s_\mathrm{in}(t) - \left(\frac{1+\Gamma_\mathrm{shunt}}{2}\right)\sum_{j=1}^4\sqrt{\kappa_{p,j}}\,p_{j}(t) \mathrm{.}
    \label{eq:time dependence output field}
\end{equation}
As a result, the output-field separation $S(t)\equiv \left|s_\mathrm{out}^e(t) - s_\mathrm{out}^g(t)\right|$ for a readout drive targeting qubit $j$ is to a good approximation given by
\begin{equation}
    S(t) = \left | \frac{1+\Gamma_\mathrm{shunt}}{2}\right  | \sqrt{\kappa_{p,j}}\,\left|p_{j}^e(t) - p_{j}^g(t) \right| \mathrm{.}
    \label{eq:time dependent output field separation}
\end{equation}
The predicted output-field separation using this expression is plotted in Fig.~\ref{fig:fast_readout_experiment}(a) of the main text for a readout drive targeting qubit Q$_2$.
\subsection{Normal modes}
\indent In the absence of an external drive, the eigenvalues ${\tilde{\lambda}_k~(k=1, \ldots, 8)}$ of Eq.~\eqref{eq:coupled_resonators_first_order_ode} give the normal modes of the coupled readout and filter resonators. Here, we set $\Gamma_{\mathrm{shunt}}=1$, as this is a good approximation around the readout- and filter-resonator frequencies and it simplifies finding the eigenvalues. The frequencies and effective external linewidths of the normal modes are then given by
\begin{equation}
\tilde{\omega}_k= \textrm{Re}\left(\tilde{\lambda}_k\right) \textrm{,} \quad 
\tilde{\kappa}_k = -2\textrm{Im}\left(\tilde{\lambda}_k\right)  \textrm{.}
\end{equation}
\begin{table}
\centering
\caption{\label{tab:Bare readout parameters - appendix} Bare readout parameters. Here, $g$ is the coupling strength between the qubit and the readout resonator, and $n_{\mathrm{crit}}$ is the critical photon number of the readout resonator.}
\begin{ruledtabular}
\begin{tabular}{cccccccc}
& $\omega_r^g/2\pi$ & $\omega_p/2\pi$ & $J/2\pi$ & $\kappa_p/2\pi$ & $\chi/2\pi$ & $g/2\pi$ & $n_{\textrm{crit}}$ \\ 
& ($\si{\mega\hertz}$) & ($\si{\mega\hertz}$) & ($\si{\mega\hertz}$) & ($\si{\mega\hertz}$) & ($\si{\mega\hertz}$) & ($\si{\mega\hertz}$) &  \\
\hline
Q$_1$ & $10250$ & $10232$ & $36.1$ & $97.6$ & $-9.4$ & $420$ & $7.0$ \\
Q$_2$ & $10386$ & $10407$ & $39.4$ & $81.4$ & $-9.9$ & $423$ & $6.7$  \\
Q$_3$ & $10540$ & $10566$ & $30.9$ & $66.7$ & $-10.5$ & $280$ & $7.1$ \\
Q$_4$ & $10666$ & $10710$ & $26.2$ & $93.5$ & $-8.3$ & $275$ & $9.4$ \\
\end{tabular}
\end{ruledtabular}
\end{table}
The eight normal modes are separated into pairs, each corresponding to the hybridized modes of a particular readout-resonator and filter-resonator pair. This picture is a slight simplification of the true nature of these modes, since in reality the coupling between the filter resonators means the normal modes hybridize across the whole network of resonators. However, by looking at the qubit-state dependence it is clear that each normal mode can be associated with a particular readout-resonator and filter-resonator pair. The dispersive shifts due to a target qubit were found by solving the normal modes for the $\ket{g}$ and $\ket{e}$ states of the target qubit with all other qubits in the ground state and using the standard definition
$\tilde{\chi}_{r} \equiv (\tilde{\omega}_{r}^e - \tilde{\omega}_{r}^g)/2$.
The dispersive shift induced in spectator normal modes (mediated via the coupling of the filter resonators) was in all cases less than $1\%$ of the shift in the target normal mode.
\\
\indent The normal modes are summarized in Table~\ref{tab:Readout system eigenmodes}. Here, $\tilde{\omega}_r^{g(e)}$ denotes the readout-resonator-like modes and $\tilde{\omega}_p^{g(e)}$ the filter-resonator-like modes. By definition, the readout-resonator-like modes have smaller external linewidths and larger dispersive shifts than the corresponding filter-resonator-like modes. For the readout characterization, measurement was performed by driving the readout-resonator-like modes, as this resulted in superior SNR. As a result, we refer to these modes as the readout modes and $\tilde{\omega}_r^g$, $\tilde{\kappa}_r^g$, $\tilde{\chi}_r$ in Table~\ref{tab:Readout system eigenmodes} correspond to $\omega_{\textrm{ro}}^g$, $\kappa_{\textrm{ro}}^g$, $\chi_{\textrm{ro}}$ in Table~\ref{tab:Readout parameters} of the main text.

%% file: appendix_body/Supplementary_F.tex
\section{ac-Stark-shift spectroscopy measurement}
\label{appendix:Ac Stark shift spectroscopy measurement}
In order to calibrate the drive power incident on the device we followed the protocol in Ref.~\cite{swiadek2023enhancing}. We first applied a continuous drive tone to the readout resonator of a particular qubit, in order to induce an ac Stark shift. We subsequently applied a $\pi$-pulse to the qubit followed by qubit measurement. The experiment was then repeated while sweeping the frequency $\omega_d$ of the $\pi$-pulse. The value of $\omega_d$ which results in the maximum qubit population exchange from the $\ket{g}$ to the $\ket{e}$ state is associated with the ac-Stark-shifted qubit frequency. The steady-state average photon number $n_{r,j}^g$ in readout resonator $j$ is then inferred from the ac Stark shift $\Delta_{\textrm{ac}}$ through the relation $\Delta_{\textrm{ac}}=2\chi_j n_{r,j}^g$~\cite{Gambetta2006}. This photon number is related to the readout-resonator coherent field amplitude $r_j^g$ by the expression $n_{r,j}^g=|r_j^g|^2$. The equations of motion Eqs.~\eqref{eq:filter mode dynamics} and~\eqref{eq:readout mode dynamics} are solved in the steady state to find the input field on the filter resonator $p_{\textrm{in},j}$ as a function of $r_g$. Finally, the incident field on the device is determined from the incident field on the filter resonator using the relation
\begin{table}
\centering
\caption{\label{tab:Readout system eigenmodes}Normal modes of the composite readout-resonator and filter-resonator system. The terms $\tilde{\chi}_{r(p)}$ are the dispersive shifts of the normal modes defined $\tilde{\omega}_{r(p)}^e \equiv \tilde{\omega}_{r(p)}^g+2\tilde{\chi}_{r(p)}$.}
\begin{ruledtabular}
\begin{tabular}{cccccccc}
& $\tilde{\omega}_{r}^g/2\pi$ &  $\tilde{\omega}_{p}^g/2\pi$ & $\tilde{\kappa}_{r}^g/2\pi$ & $\tilde{\kappa}_{r}^e/2\pi$ & $\tilde{\kappa}_{p}^g/2\pi$ & $\tilde{\chi}_{r}/2\pi$ & $\tilde{\chi}_{p}/2\pi$\\ 
 & (\si{\mega\hertz}) & (\si{\mega\hertz}) &  (\si{\mega\hertz}) & (\si{\mega\hertz}) & (\si{\mega\hertz}) & (\si{\mega\hertz}) & (\si{\mega\hertz}) \\ \hline
Q$_1$ & $10221$ & $10284$ &  $42$ & $30$ & $42$  & $-5.9$ & $-3.5$ \\
Q$_2$ & $10360$ & $10438$ &  $34$ & $25$ & $63$  & $-7.8$ & $-2.3$ \\
Q$_3$ & $10520$ & $10582$ & $24$ & $14$ & $55$  & $-8.4$ & $-2.3$ \\
Q$_4$ & $10652$ & $10701$ & $19$ & $11$ & $59$  & $-7.2$ & $-1.1$ \\
\end{tabular}
\end{ruledtabular}
\end{table}
\begin{equation}
    \frac{s_{\textrm{in}}}{p_{\textrm{in},j}} = \frac{1 + \Gamma_{p,j}}{1+\Gamma_{\textrm{incident}}} \textrm{.} 
\end{equation}
Here, $\Gamma_{\textrm{incident}}$ and $\Gamma_{p,j}$ are the reflection coefficients defined in Appendix~\ref{appendix:Semi-classical model for the readout system dynamics}, which are determined from the readout and shunt parameters. The squared norm of the incident field $|s_{\textrm{in}}|^2$ corresponds to the photon flux $\dot{n}$ on the device, and thus the power incident on the device is given by $P=\hbar\omega_d|s_{\textrm{in}}|^2$.
\begin{figure}
\includegraphics[width=0.475\textwidth,height=0.475\textheight,keepaspectratio]{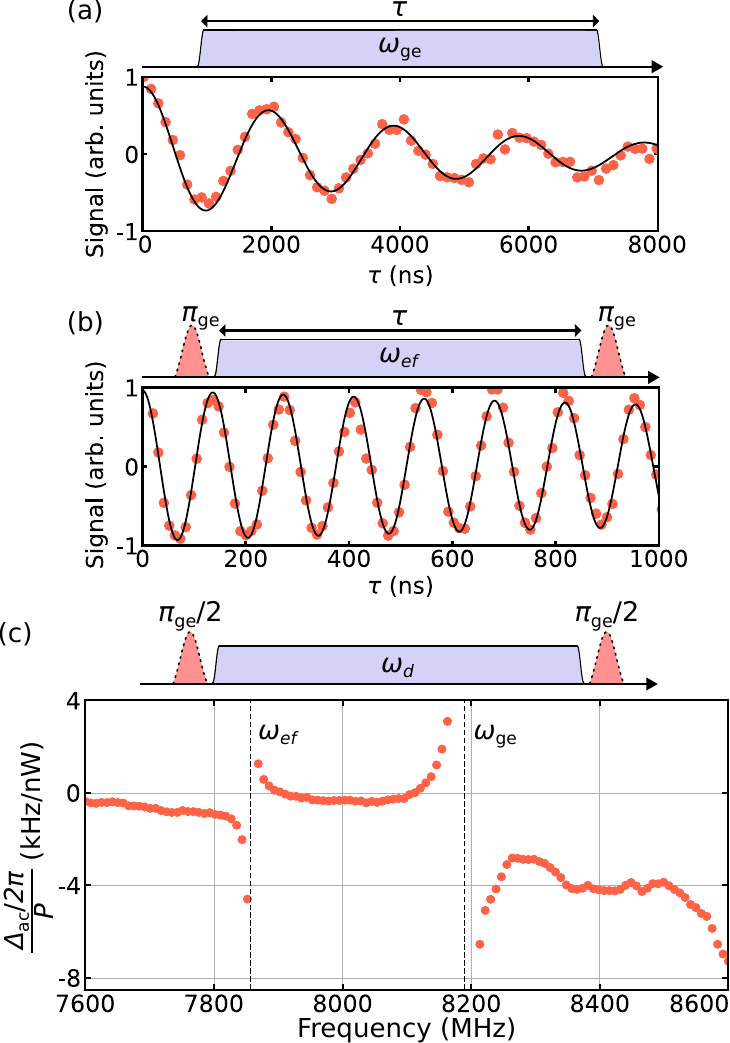}
\caption{\label{fig:Purcell_filtering_measurements - appendix} (a)--(b)~Rabi oscillations in qubit Q$_1$ driven at the $|g\rangle$--$|e\rangle$ and the $|e\rangle$--$|f\rangle$ transition frequencies, respectively. (c)~Pulse sequence for measuring the ac Stark shift and results for qubit Q$_2$. The ratio between the Stark shift and incident power $P$ as a function of drive frequency.}
\end{figure}
\\
\indent From the calibrated power incident on the device, we could determine the attenuation between the generator and the device. We make the assumption that this attenuation is the same over the range of frequencies that are driven in the subsequent section. Since the attenuation of the coaxial cables of the input lines increases with frequency, this assumption will lead to lower-bound values for the Purcell-limited relaxation time.
\\
\indent Next, the drive amplitude $\Omega$ incident on the qubit was determined as a function of frequency. Here, $\Omega$ is defined in terms of the drive Hamiltonian as $\hatH_{\textrm{drive}}=\hbar\Omega\left(a_q^\dag + a_q\right)\cos \! \left(\omega_d t\right)$ where $a_q^\dag$ and $a_q$ are the creation and annihilation operators for the qubit.
\begin{figure*}
\includegraphics[width=0.99\textwidth,height=0.99\textheight,keepaspectratio]{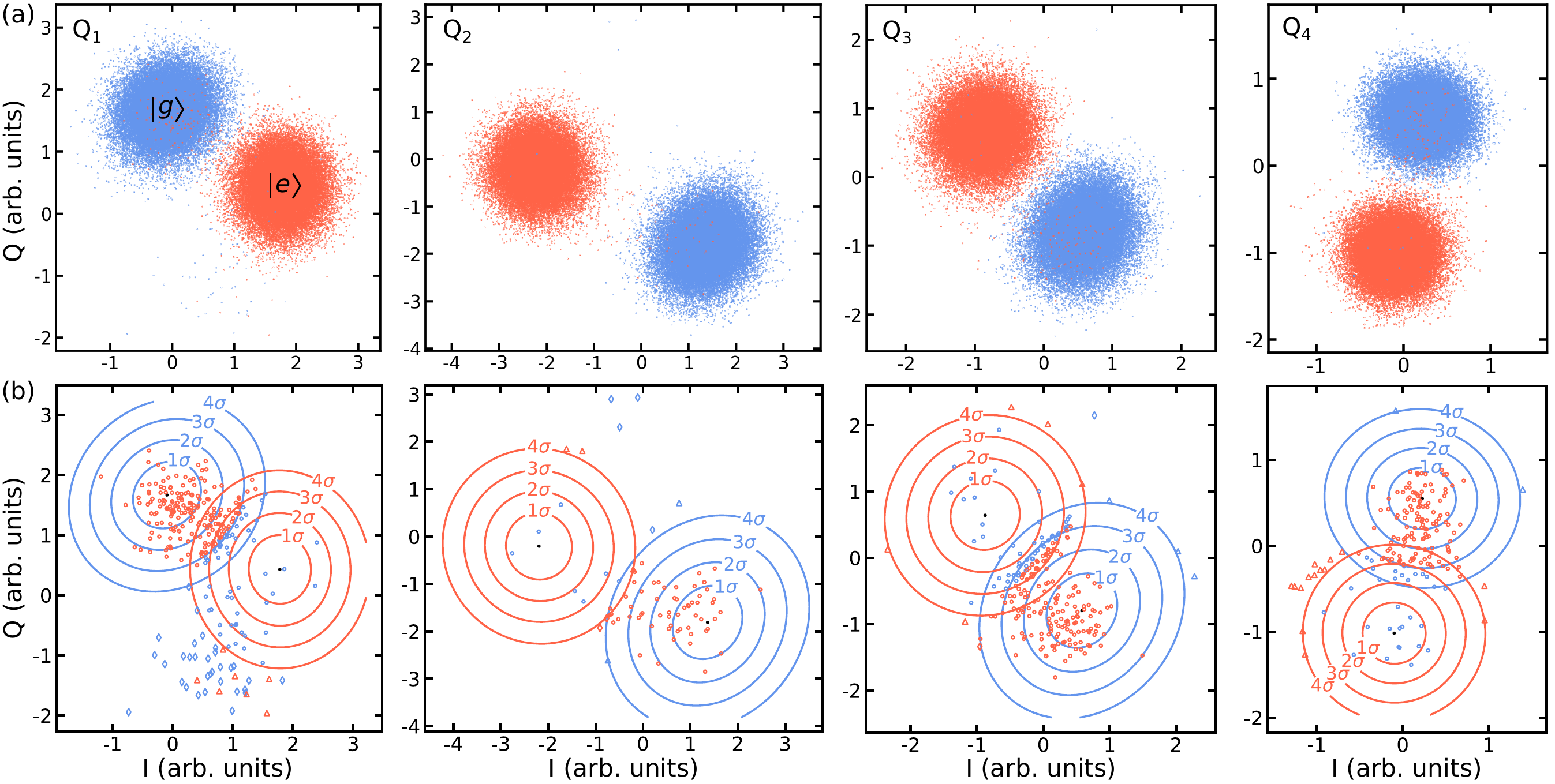}
\caption{\label{fig:IQ_single_shot_analysis} Post-selected IQ signals from the assignment-fidelity pulse sequence. (a) Time-integrated IQ signals when the qubit was prepared in the $\ket{g}$ state (blue) and in the $\ket{e}$ state (red). (b) The $1\sigma$ to $4\sigma$ confidence ellipses of the fitted bivariate normal distributions. Misassigned shots that lie inside (outside) both of the $4\sigma$ ellipses are shown as circles (diamonds). Correctly assigned shots that lie outside the $4\sigma$ ellipse of the assigned state are shown as triangles.}
\end{figure*}
\\
\indent As described in the main text, $\Omega$ was determined by two methods following Ref.~\citenum{sunada2022fast}: (i) by measuring the Rabi frequencies of the qubit when driving at the $|g\rangle$--$|e\rangle$ and $|e\rangle$--$|f\rangle$ transition frequencies, and (ii) by measuring the induced ac Stark shift $\Delta_{\mathrm{ac}}$ in the qubit when detuned from these transitions. The drive is applied through the readout line. Figures~\ref{fig:Purcell_filtering_measurements - appendix}(a) and~(b) show Rabi oscillations in Q$_1$. The $|g\rangle$--$|e\rangle$ and $|e\rangle$--$|f\rangle$ Rabi oscillation frequencies are equal to $\Omega$ and $\sqrt{2}\,\Omega$, respectively.
\\
\indent Figure~\ref{fig:Purcell_filtering_measurements - appendix}(c) shows the measured ac Stark shift $\Delta_{\mathrm{ac}}$ normalized by the incident power for qubit Q$_2$ as a function of the drive frequency. The drive amplitude $\Omega$ was then determined from the measured ac Stark shift using the perturbative formula in Ref.~\citenum{sunada2022fast}. The results are sensitive to the calibrated value of the qubit frequency $\omega_q$. We determined this more precisely by applying a Stark shifting drive to the qubit over a range of powers $P$, and performing a linear fit of the Stark-shifted qubit frequency $\omega_{q,\mathrm{ac}}$ to the formula $\omega_{q,\mathrm{ac}} = \omega_q+kP$, where $k$ is the slope of the linear fit. From the fit, we extracted the value of $\omega_q$ and an associated uncertainty. \\
\indent In qubits Q$_1$, Q$_2$ and Q$_3$, we observed a frequency range between the $|g\rangle$--$|e\rangle$ and $|e\rangle$--$|f\rangle$ transition frequencies where the drive induced a small negative ac Stark shift in the qubit---contrary to the positive shift that is expected across this frequency range. This may result from spurious harmonics of the generator producing the Stark shifting drive, since the sensitivity to the main drive tone is strongly suppressed around the notch frequency. We excluded the data in this range when determining the Purcell-limited relaxation time.
\\
\indent From the measured drive amplitude and the incident power, the Purcell-limited relaxation time of the qubit is given by~\cite{sunada2022fast}
\begin{equation}
    T_1^\mathrm{pl}(\omega_d)=\frac{4P}{\Omega^2 \hbar\omega_d} \textrm{.}
\end{equation}


%% file: appendix_body/Supplementary_G.tex
\section{Further analysis of the fast multiplexed qubit readout}
\label{appendix: Readout analysis}
\begin{table}
\caption{\label{tab:Readout further info - appendix} Further information regarding the readout. Here, $\sigma_e/\sigma_g$ expresses the ratio of the readout signal standard deviations for the $\ket{g}$ and $\ket{e}$ states. The terms $P_{0}(e_2|g_1)$ and $P_{\pi}(g_2|g_1)$ are the error contributions to the assignment fidelity when the qubit was prepared in the $\ket{g}$ and $\ket{e}$ states, respectively. The term $P_{e}$ is the excited-state population of the qubits prior to state-preparation, and $\omega_d$ is the drive frequency that was used to perform readout.}
\begin{ruledtabular}
\begin{tabular}{ccccccc}
& SNR & $\sigma_e/\sigma_g$ & $P_{0}(e_2|g_1)$  & $P_\pi(g_2|g_1)$ &$P_{e}$ &  $\omega_d/2\pi$ \\ 
  &  & & (\si{\percent}) &(\si{\percent}) & (\si{\percent}) & (\si{\mega\hertz}) \\
\hline
Q$_1$ & $6.3$ & $1.03$ & $0.22$ & $0.49$ & $1.7$ &  $10224$ \\
Q$_2$ & $8.4$ & $0.99$ & $0.03$ & $0.16$ & $2.3$ & $10357$\\
Q$_3$ & $6.0$ & $1.03$ & $0.13$ & $0.44$ & $1.4$ & $10515$\\
Q$_4$ & $6.7$ & $0.99$ & $0.08$ & $0.33$ & $1.0$ & $10646$\\
\end{tabular}
\end{ruledtabular}
\end{table}
Here, further details of the fast multiplexed readout measurement are provided. First, the calculation of the separation error and coherence limited errors is discussed. The separation error was determined from the signal-to-noise ratio (SNR) of the measurements. The SNR is here defined as
\begin{equation}
\textrm{SNR} \equiv \frac{\left\vert\mu_g - \mu_e \right\vert}{(\sigma_g + \sigma_e)/2}  \textrm{,}
\end{equation}
where $\mu_{g(e)}$ and $\sigma_{g(e)}$ are the the mean and standard deviation, respectively, of the Gaussian fits to the $\ket{g(e)}$ state histograms shown in Fig.~\ref{fig:fast_readout_experiment}(c) of the main text. The measured SNR values for the four qubits are provided in Table~\ref{tab:Readout further info - appendix}. The ratio $\sigma_e/\sigma_g$ is also given, showing that the noise for the $\ket{g}$ and $\ket{e}$ states was almost the same. The separation error due to overlap of the projected Gaussian distributions is estimated from the SNR as~\cite{swiadek2023enhancing, gambetta2007protocols}
\begin{equation}
\varepsilon_{\textrm{sep}} = \frac{1}{2}\left[1 - \textrm{erf}\left(\frac{\textrm{SNR}}{\sqrt{8}}\right)\right] \textrm{.}
\end{equation}
The coherence limit to the assignment fidelity is estimated as
\begin{equation}
\varepsilon_{\textrm{cl}} =  \frac{\tau_{\textrm{meas}}}{2T_1} \textrm{,}
\end{equation}
where $\tau_{\textrm{meas}}$ is the length of the measurement integration window. The coherence limit to the QND fidelity is estimated as
\begin{equation}
\varepsilon_{\textrm{cl}}^Q =  \frac{\tau_{\textrm{buffer}}}{2T_1} + \frac{\tau_{\textrm{meas}}}{{2T_1}} \textrm{,}
\end{equation}
where $\tau_{\textrm{buffer}}$ is the length of the buffer period between the first and second measurements of the QND measurement pulse sequence. 
\\
\indent Figure~\ref{fig:IQ_single_shot_analysis} shows the IQ plots for the simultaneous measurement on the four qubits. These measurements are post-selected to ensure the qubits were prepared in the $\ket{g}$ state prior to the $\pi$ pulse sequence. Table~\ref{tab:Readout further info - appendix} gives the fraction of discarded measurements, represented by the excited-state population $P_{e}$. In Fig.~\ref{fig:IQ_single_shot_analysis}(a), all of the $\sim$$8\times10^4$ post-selected shots are plotted. We fit the measured $|g\rangle$- and $|e\rangle$-state data to bivariate normal distributions. From the fits, we determined the confidence ellipses. Figure~\ref{fig:IQ_single_shot_analysis}(b) shows the $1\sigma$ ($68.27\%$), $2\sigma$ ($95.45\%$), $3\sigma$ ($99.73\%$) and $4\sigma$ ($99.994\%$) ellipses. We stress that this $\sigma$ is not referring to a standard deviation. Rather, the interpretation is that $68.27\%$ of single-shot measurements are expected to lie inside the $1\sigma$ confidence ellipse, $95.45\%$ inside the $2\sigma$ ellipse, etc. Shots that were misassigned are highlighted. These are divided into two categories. Shots that were misassigned and that lie inside (outside) the $4\sigma$ ellipses for both the $|g\rangle$ and $|e\rangle$ states are shown as circles (diamonds). We attribute misassigned shots outside the $4\sigma$ ellipses to measurement-induced excitation to leakage states as we would not expect these events if the qubit were restricted to the $|g\rangle$--$|e\rangle$ subspace. In addition, correctly assigned shots that lie outside the $4\sigma$ ellipse of the assigned state are shown as triangles. We expect approximately 5 of these events across $8\times10^4$ measurements. When many of these events cluster together, we again attribute it to measurement-induced leakage. \\
\indent For Q$_1$, there is a clear signature of leakage from the $|g\rangle$ state that made a significant contribution to the assignment error. There is also a signature of leakage from the $|e\rangle$ state that did not lead to assignment errors. In the remaining qubits, $>$$4\sigma$ measurements with signatures of leakage affect less than $0.015\%$ of the total measurements. However, we emphasize that the absence of high-$\sigma$ events in the IQ plane does not preclude all leakage events. For instance, it is possible for the output field given the qubit is in a leakage state to overlap with the $|g\rangle$- or $|e\rangle$-state output fields. Further characterization such as pseudo-syndrome detection~\cite{Hazra2024benchmarking} could be performed in order to detect such `hidden' leakage events.